# Investor Sentiment in Asset Pricing Models: A Review of Empirical Evidence

**Szymon Lis, University of Warsaw, Corresponding author: sm.lis@student.uw.edu.pl**

**Abstract**

This study conducted a comprehensive review of 71 papers published between 2000 and 2021 that employed various measures of investor sentiment to model returns. The analysis indicates that higher complexity of sentiment measures and models improves the coefficient of determination. However, there was insufficient evidence to support that models incorporating more complex sentiment measures have better predictive power than those employing simpler proxies. Additionally, the significance of sentiment varies based on the asset and time period being analyzed, suggesting that the consensus relying on the BW index as a sentiment measure may be subject to change.

## 1. Introduction

Asset pricing models have been extensively studied by economists over several decades, with the Capital Asset Pricing Model (CAPM) being the most basic and widely used model to determine the expected rate of return (Sharpe, 1964). Fama and French (1992) enhanced the CAPM by incorporating company size and book-to-market (BM) value, creating the three-factor model, which has since been expanded upon by several other models, including Jegadeesh and Titman (1993), Carhart (1997), Fama and French (2013), Hou et al. (2015), and Fama and French (2015). These models follow a rational paradigm, which assumes that competition among rational investors leads to an equilibrium where prices equal the rationally discounted value of expected cash flows (Friedman, 1953; Fama, 1965).

However, some market participants do not follow a rational paradigm. They are known as "noise traders". They buy or sell based on not technical or fundamental analysis, but hype/rumors called investor sentiment and in fact, it can heavily affect prices. The definition of sentiment is imprecise. For this study, Baker and Wurgler's (2007, p. 129) definition of investor sentiment is used: "a belief about future cash flows and investment risks that is not justified by the facts at hand". This can be extended to the overall attitude or feeling of investors about a particular stock, market, or economic situation. It can be positive, neutral, or negative, and it can change rapidly based on a variety of factors, such as economic data, company announcements, and global events.

Various theories on investor sentiment and noise traders exist, but the most widely applied behavioral theory is De Long et al.'s (1990), which predicts that investor sentiment can impact stock returns and persist in financial markets, causing asset prices to deviate from fundamental values and leading to inefficient allocation of capital. This has significant implications for investors' portfolio allocation decisions, firms' cost of capital, and even the decision-making process of central banks and government agencies (Smales, 2017). Therefore, understanding investor sentiment, identifying appropriate sentiment measures, and quantifying its impact on asset prices are essential topics in finance research.

As a result, there has been a growing interest among researchers to incorporate behavioral factors, including sentiment, into asset pricing models due to the discovery of anomalies (Huang et al., 2016). The momentum effect was first demonstrated by Jegadeesh and Titman (1993), and DeBondt and Thaler (1985, 1987) found evidence of long-term reversal over 3-5 years. Subsequently, numerous empirical studies attempted to measure investor sentiment (Lee et al., 1991; Brown and Cliff, 2004; Fisher and Statman, 2000; Brown and Cliff, 2005), showing that individual investors can be easily swayed by sentiment. Moreover, sentiment indicators can increase the explanatory power of traditional models for stock returns that are difficult to arbitrage and value, such as small stocks, value stocks, low-priced stocks, and stocks with low institutional ownerships. Therefore, incorporating investor sentiment into asset pricing models can improve our understanding of market anomalies and help investors make better-informed decisions.

The motivation behind this study is to address the significant yet complex role of investor sentiment in asset pricing models. Despite the number of published works on the issue of investor sentiment, the results did not allow us to obtain coherent knowledge about sentiment as most of the researchers used different measures and various models to study the impact of sentiment on stock returns. In particular, the empirical question of a relationship between investor sentiment and stock market behavior remains unclear. Some review articles tried to provide a synthesis of the behavioral finance literature. However, they were not focused on sentiment itself (i.e. sentiment was only the part of the analysis for which several articles were analyzed), and now they are also relatively obsolete (among others Subrahmanyam, 2007). Hence, given the important role of investor sentiment in asset pricing and that so much has been written on the subject, this paper aims to provide comprehensive coverage of the status of this research. Taking a utilitarian viewpoint, the success of an asset pricing model lies in its explanatory and out-of-sample forecasting power. However, it is impossible, in practice, to perform tests on all asset pricing models on many data sets and over many different periods. This paper aims to provide a comprehensive overview of the literature on the application of investor sentiment in asset pricing models. Therefore, this study focuses on reviewing the methodologies and empirical findings of 71 papers that apply different investor sentiment measures for modeling stocks and indices' returns in the markets around the world (for details see section *3.1 The scope of the study*).

The contribution of this review is to provide a bird's-eye view of the whole return forecasting literature and to provide some recommendations for the practice and future research. Therefore, the study search for an answer to the following main research question (RQ):

*What is the impact of investor sentiment on stocks and indices returns in the presence of other market factors?*

To answer that question based on the available literature, two auxiliary hypotheses were formulated:

*RH1. Augmenting models with the investor sentiment proxies improves the coefficient of determination.*

*RH2. Models with more complex sentiment have better predictive power than those with simpler proxies or those using only individual measures.*

Some formulations from the above hypotheses require clarification:

1) The term 'models', as used here, refers to all models employed in the analyzed papers in a general sense. This includes single-factor, medium complex models that incorporate various factors, excluding those present in multifactor models such as macroeconomic variables, as well as multifactor and Machine Learning (ML) models.
2) 'Complex sentiment' means a sentiment created based on at least two individual sentiment measures or sentiment based on data from social or mass media. Simple measures are single sentiment indicators, e.g. survey results, returns on IPOs, or Google Search Volume.

The remaining sections are organized as follows. Section 2 provides some preliminaries such as the definition and measurement of investor sentiment and explains the model used in research. Section 3 introduces the methodology and materials used in this study, i.e. method of selection and analysis of articles. Section 4 provides the results of the study. Section 5 summarizes and concludes. Appendix A contains a summary of each of the 71 papers.

## 2. Some Preliminaries

*2.1 Sentiment measures*

2.1.1. Sentiment definition

The sentiment does not have an indisputable definition (Zhang, 2008). Existing definitions of sentiment in the literature range from vague statements about investors' mistakes to various psychological biases (Shefrin and Belotti, 2007) e.g. general investor attitudes towards markets (Shleifer and Summers, 1990), investor optimism or pessimism (Antoniou et al., 2013) or beliefs about equity returns (Barberis et al., 1998). Furthermore, the term itself is subject to a wide spectrum of classifications and is used in different ways by academic researchers, financial analysts, and the media (Barberis et al., 1998; Daniel et al., 1998; Welch and Qiu, 2004; Cliff and Brown, 2004; Shefrin and Belotti, 2007; Baker and Wurgler, 2007). Due to the lack of one consistent definition of sentiment, it is also impossible to present one formula that would sufficiently illustrate or optionally (with appropriate data) allow to calculate of the value of the sentiment. However, on some level, it is possible (Zhou, 2018). If sentiment relates only to the over- or under-valuation of assets, the sentiment $S_t$ can be defined as the difference between the price observed in the market $P_t$ and the fundamental price estimated from a rational benchmark asset pricing model $P_t^*$:

$$S_t = P_t - P_t^*  \quad (1)$$

With the above definition, $S_t = 0$ implies that the market price agrees with the fundamental value. In practice, however, $S_t$ is rarely zero. The greater the $S_t$, the more optimistic investors are about the asset value. $S_t$ can of course be negative, representing pessimism about the asset value. Likewise, investor sentiment can be also derived based on returns:

$$S_t = r_t - r_t^*  \qquad (2)$$

where $r_t$ is the observed or expected return and $r_t^*$ is the fundamental return from a rational benchmark asset pricing model. In general, we can define sentiment in terms of any characteristic (CH):

$$S_t = CH_t - CH_t^*  \qquad (3)$$

Existing sentiment studies largely rely on prices, returns, and expected probabilities; Much work remains to be done on volatility sentiment, tail sentiment, and sentiment of other characteristics of asset returns. Most research focused on developing indicators based on proxies without a well-defined sentiment. The common point is the reliance on information for which there is no fundamental foundation.

*2.1.2 Indirect and direct sentiment*

While there is no single definition of sentiment, researchers agree on two types of investor sentiment measures. The research distinguishes direct and indirect measures of sentiment. However, this distinction is not clear. For this study, the assumption is made that direct measures are derived from surveys that capture opinions on the stock market conditions, such as the Michigan Consumer Sentiment Index (MCSI) or Investors' Intelligence (II) (Qiu and Welch, 2004). Whereas indirect ones represent economic variables such as the closed-end fund discount or returns on IPOs that capture an investor's state of mind. The second category also includes data available from sources like Natural Language Processing (NLP) for social media or news, and exogenous non-economic factors e.g. cloud cover (Hirshleifer and Shumway, 2003). Both types of measures have their advantages and disadvantages. For modeling purposes, indirect measures have an advantage over direct ones. They are easy to construct, observed in real-time, and reflect both rises and falls in the market. However, they are difficult to interpret and some of them are based on questionable theoretical fundamentals. Moreover, these indicators are a combination of expectations and sentiment. The process of separating one from the other may be difficult, if not impossible (Beer and Zouaoui, 2013). But this argument refers also to direct ones. Also, for this study, it was assumed that the combined measure of sentiment comes from at least two data sources.

*2.1.3 Theories of investor sentiment*

The concept of market sentiment is not a new idea and was noticed even among the most popular economists, e.g. Keynes (1936) provided an early analysis of speculative markets and investor sentiment. However, even now, there is no consensus on the theoretical structure of behavioral finance and the investor sentiment research area (Ángeles López-Cabarcos et al., 2020).

The most popular in terms of citations is the prospect theory created by Kahneman and Tversky (1979). This theory is an alternative to the expected utility theory. Its authors showed that the value assigned to gains or losses is assessed asymmetrically. That fact may significantly affect equity prices. Tversky and Kahneman (1974) also explained three other heuristics, i.e. representativeness, availability of instances or scenarios, and adjustment from an anchor. They employed them in making judgments under uncertainty and showed their application in economics. De Bondt and Thaler (1985) analyzed how over-reaction behavior influences stock prices, finding inefficiencies in the weak form market proposed by the efficient market's theory. Shiller et al. (1984) explained sentiment in terms of social dynamics. Shefrin and Statman (1985) dealt with the behavior pattern known as the disposition effect, which causes investors to sell winners too early and hold losers for too long. Winners are those assets that in the previous period had a positive rate of return, while losers are those that brought losses. Daniel et al. (2005) analyzed two psychological biases experienced by an investor, i.e. overconfidence and biased self-attribution to the securities market. They showed that both biases affect volatility, short-run earnings, and future returns. Barberis et al. (1998) proposed a model of investor sentiment. It is based on psychological evidence regarding conservatism, and the representativeness heuristic, producing under- and over-reaction and showing that investor sentiment is related to these behaviors.

However, the most influential work regarding investor sentiment was presented by De Long et al. (1990). They presented a model whereby irrational noise traders affect prices. The existence of noise traders was theoretically accepted as a solution to the results achieved by Grossman and Stiglitz (1980) which showed that under most circumstances an investor with superior information cannot get a higher profit based on that information. The theory explains some anomalies such as the excess volatility of asset prices or the mean reversion of stock returns and provides evidence that assets exposed to the noise traders are riskier and offer a return premium. As emphasized by Cochrane (2008), the market risk premium has important implications in all areas of finance e.g. it can lead to market bubbles followed by massive devaluations (Brown and Cliff, 2004) or it enables to create of profitable trading strategies (Baker and Wurgler, 2006; Fisher and Statman, 2000). Subsequently, several models have been proposed to explain sentiment. Thaler (1993) and Brunnermeier (2001) reviewed some of the first advances. Shefrin (2008) provided a synthesis of behavioral theories in the stochastic discount framework of asset pricing. Baker and Wurgler (2012) reviewed the implications of sentiment models in the context of corporate decision-making. In recent work, Greenwood et al. (2016) used extrapolation learning to explain credit sentiment. Some scholars have proposed psychological and behavioral decision theories to explain many abnormal effects, including overreaction, under-reaction, overconfidence, group behavior, the emergence of speculative bubbles, the excessive volatility of the stock market, and so on (Barberis and Thaler, 2003). Many works challenge the models presented above (among others Loewenstein and Willard, 2006; Fama, 2021).

The size of behavioral theoretical schools makes it difficult to interpret the results, hence causing the research is not consistent with each other and slowing down the creation of consistent knowledge among researchers. However, this situation did not arise without a reason. High investor sentiment can mean investors are bullish about stock markets (Liu, 2015), which can produce noise trading (De Long et al., 1990; Renault, 2017). High investor sentiment could also indicate high overconfidence (Odean, 1998). Additionally, investor sentiment can be related to over-reaction, which increases when investor sentiment is low (Piccoli and Chaudhury, 2018). Notwithstanding undoubtedly, it is due to such differences between theories and empirical results that this field has recently gained a lot of interest.

*2.1.4 Measures of sentiment*

Starting with the direct measures of investor sentiment. The American Association of Individual Investors (AAII) conducts a monthly allocation survey since 1987 that asks participants whether they have a bullish or bearish attitude toward the market. De Bondt (1993) found that individual investors surveyed by the AAII forecast future stock returns. Solt and Statman (1988) and Clarke and Statman (1998) point out that investor sentiment compiled by the II survey is not useful as an indicator. The second popular survey measure, i.e. the MCSI Index focuses on five questions related to perceiving future financial situation by respondents. Answers are coded on a scale from 1 (good) to 5 (bad) and averaged (equal-weighted). Lemmon and Portniaguina (2006) find that the sentiment index as proxied by consumer confidence can forecast the returns of small stocks and those with low institutional ownership.

In terms of direct measures, there is a plenty of them from various sources. The most popular is the BW index created by Baker and Wurgler (2006) as a linear combination of popular indirect measures, i.e. closed-end fund discount, stock turnover ratio, number of Initial Public Offerings (IPOs), average returns on the day per IPO, the share of equity issue in a total issuance of shares and debt (EQTI), and dividend premium. The study showed that measures combining a lot of information about sentiment are better at predicting than single measures. A wider comparison of all available measures indicated that complex sentiment indices (both those consisting only of indirect measures and those that also contain direct ones) dominated in forecasting returns. Single direct indicators predicted only a size premium and from individual indirect proxies, only the ratio of odd purchases to sales predicted portfolios' returns sorted by size, profitability, and tangibility (Beer and Zouaoui, 2013). Now combined proxies are one of the most popular applied for asset pricing purposes. There are several influential studies on the measure of media-based sentiment (Narayan and Bannigidadmath, 2017; Narayan, 2020; Li et al., 2020). For example, Narayan et al. (2017a, 2017b) and Narayan (2019) used financial news to reflect public expectations for stock returns. In addition to the use of news to measure sentiment, indicators derived from social media have attracted more and more attention, since investors not only read market information but also share individual investment opinions on social media (Oliveira et al., 2016).

*2.1.5 Alternative indicators*

Some measures are unusual as weather variables, media-based proxies (based on Twitter, Google, or news data), or even some indicators which have never been used in the study of asset pricing. These could be investor sentiment measures that can be subscribed to serve as a proxy for psychological bias, i.e. herd behavior (Cipriani and Guarino, 2014). One of the popular examples of such an indicator is the Economic Policy Uncertainty (EPU) Index, developed by Baker et al. (2016). The index has three basic components. The first is based on a normalized measure of the volume of news articles discussing EPU from 10 newspapers in the U.S. The second component relies on reports by The Congressional Budget Office, which publishes a list of temporary federal tax code provisions. The third component uses the disagreement among economic forecasters as an indicator of uncertainty. The data are published monthly and daily starting from 1985. Some researchers also developed the EPU index for other countries like EU countries (Phan et al., 2018) and China (Chen et al., 2017).

*2.2 Application of investor sentiment*

The applications of investor sentiment do not end with asset pricing models. Some research also supports the predictive effect of sentiment on volatility. At first, Mitchell and Mulherin (1994) found weak correlations between market volatility and news. Later, Antweiler and Frank (2004) investigated the number and the tone of stock market posts on Yahoo! Finance and Raging Bull for large US stocks. The number predicted the volatility of returns and the volume. Guidolin and Pedio (2021) showed that GARCH models augmented to include media coverage and media tone outperformed traditional GARCH models for FTSE 100 returns. However, so for there are no studies that test whether investor sentiment could improve forecasting down-side risk measures such as Value-at-Risk or Expected shortfall.

Some studies use investor sentiment to price commodities (He et al., 2019, Balcilar et al., 2017) and research that tests causality between indirect and direct investor sentiment proxies. Brown and Cliff (2004) analyzed various direct and indirect sentiment indicators. They showed a correlation between them. Many popular indicators are related to survey data and are significant as regressors to predict direct measures. However, some research using different data found no association between direct and indirect measures (Qiu and Welch, 2006)

*2.3 Multifactor models*

This section describes the multifactor models used in the analyzed studies, to which researchers added a sentiment measure as an exogenous variable.

*2.3.1 CAPM*

Capital Asset Pricing Model (CAPM) is a model that allows illustrates the relationship between the incurred systematic risk and the expected rate of return on a portfolio of financial assets (Sharpe, 1964). The equation of this model is as follows:

$$R_i - R_f = \alpha_i + \beta_i * (R_m - R_f) \qquad (4)$$

where:

$R_i$ – the expected rate of return on the ith portfolio of financial assets;
$R_f$ – risk-free rate;
$R_m$ – return on the market portfolio;
$\alpha_i$ – the intercept;
$\beta_i$ (beta) – the sensitivity of the expected excess of the ith asset returns to the expected excess market returns.

*2.3.2 FF three-factor model*

Fama–French (FF) three-factor model is a model designed by Fama and French in 1992. They added two factors to CAPM to reflect a portfolio's exposure to these two classes:

$$R_i - R_f = \alpha_i + \beta_i * (R_m - R_f) + \beta_{i,s} * SMB + \beta_{i,v} * HML \qquad (5)$$

where:
$SMB$ – size premium (Small Minus Big);

*HML* – value premium (High Minus Low);
$\beta_{i,s}, \beta_{i,v}$ – factor coefficients.

*2.3.3 Carhart four-factor model*

Carhart (1997) presented the model as a tool for valuating mutual funds. He based his work on Jegadeesh and Titman´s (1993) article which revealed a tendency for good and bad performances of stocks to persist over a couple of months, in other words, a momentum effect. Thus, Carhart added the WML (Winner Minus Lossers, i.e. the return of the momentum factor) factor to the FF three-factor model:

$$R_i - R_f = \alpha_i + \beta_i * (R_m - R_f) + \beta_{i,s} * SMB + \beta_{i,v} * HML + \beta_{i,m} * WML \qquad (6)$$

In addition to the models described here, in individual cases, there were other models such as the six-factor model which is Carhart's four-factor model augmented with factors for short-term and long-term reversal, and models augmented with the MGMT factor that is constructed from a set of six anomaly variables that can be directly influenced by a firm's management (Fang and Taylor, 2021) or the PERF factor constructed from five anomaly variables that represent a firm's performance (Fang and Taylor, 2021).

*2.4 Assessment of models*

Depending on the hypotheses being verified, researchers are interested in different statistics. Most often, the hypotheses concern the significance (p-value lower than 0.05 or 0.01) of the investor sentiment proxy and the expected sign of the relationship between expected return and investor sentiment. When the study is to compare multiple countries or statistics, researchers sometimes refer to adjusted R-squared or incremental adjusted R-squared, i.e. how the adjusted R-squared increases after adding a proxy of the investor sentiment.

**3. Methods and Materials**

*3.1 The scope of the study*

This study examines empirical studies that employ quantitative analyses of investor sentiment and its impact on asset pricing models across the globe, i.e. North America, Europe, and Asia. The search for relevant articles was conducted on the Web of Science (WoS) platform. It is important to acknowledge that the field of behavioral finance is extensive, and it is not feasible to review every publication. Consequently, some subjective choices regarding which scholarly articles to consider are inevitable. The articles analyzed in this study were published between 2000 and 2021, and have a minimum of 60 citations (as indicated on the WoS website in January 2022[1]). Each of the searchable fields must include at least one of the following terms: 'Sentiment indicator',

---

[1] We included only publications that were at least a year old so that the criterion of the number of citations did not exclude new important research and ensured a robust review in the period under study.

'Sentiment proxy' or 'Investor sentiment'. The initial search resulted in a sample of 232 papers. However, papers with the following characteristics were excluded from this review:

1) exclusively calculate correlations or test causation between investor sentiment and stock returns;
2) forecast a characteristic other than return (e.g. volatility);
3) employ non-empirical methodologies such as experiments or theoretical backgrounds;
4) analyze prices other than stocks, such as currencies (including cryptocurrencies), commodities, options, etc;
5) do not provide detailed data on sentiment measures, stock returns, and model specifications;
6) were published in any other language than english.

Figure 1 depicts the process of article selection for review.

**Figure 1.** Flowchart of Literature Review Process with Inclusion and Exclusion Criteria.

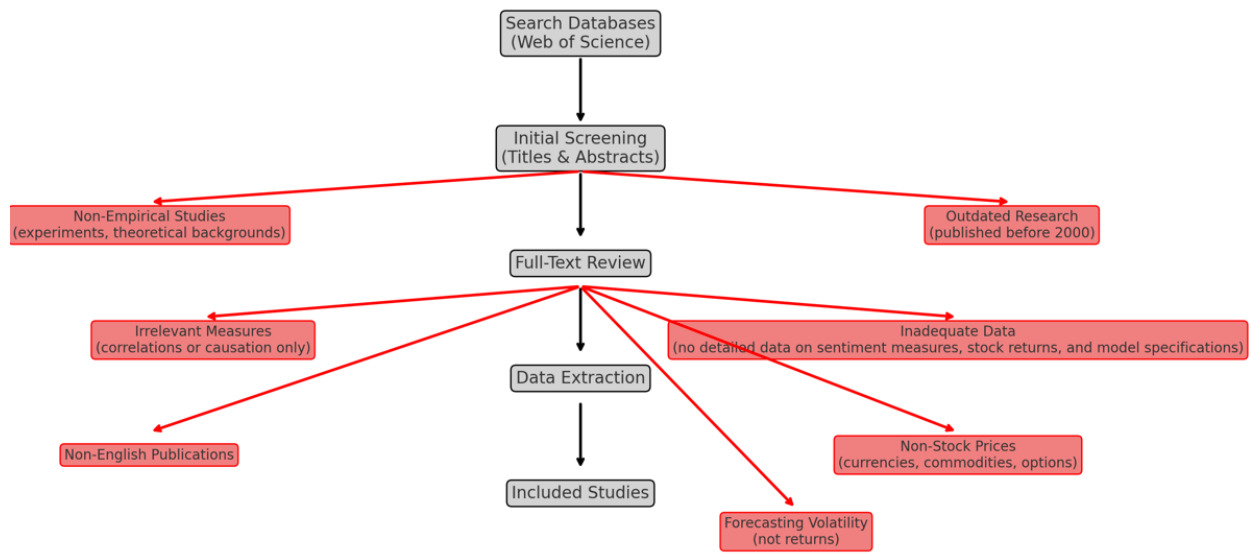

After excluding these papers, the final sample included 71 articles, which are listed in Appendix A.

*3.2 Methodology of analyzing articles*

The articles were analyzed qualitatively and quantitatively. The first type of analysis focused on getting to know the research methodology, the way of creating a measure of sentiment, the models used, descriptions and explanations of the obtained results, and its references to similar studies. For quantitative analysis, data consisted of the coefficient for sentiment and other factors, adjusted R2, as well as t-statistics and standard errors. After that weighted averages and frequencies of occurrence were calculated.

## 4. Results

In this review, the author examines various aspects of the literature on investor sentiment and asset pricing. The analysis encompasses measures of sentiment, models employed, the scope of studies, and the years under investigation. Despite the significant number of studies conducted on the topic, the evidence supporting the notion that investor sentiment plays a role in asset pricing remains inconclusive due to the limited control of other variables in most research. Furthermore, many studies apply only one or two models without comparing results between them. Additionally, while there is one popular investor sentiment measure, when a new one is introduced, it is not compared to this or any other measure. Finally, it appears that other characteristics are far more relevant in determining the cross-section of expected returns in each case.

Table 1 presents numbers and percentage shares of articles analyzed in this study divided by various characteristics of papers[2].

First, most of the studies were conducted on the U.S. stocks market, which constitutes 54 of the 71 papers analyzed. There were also four studies regarding the Chinese market, one per German, Japanese, and UK markets, and what is important in terms of generalizing the results, 11 studies concerning multiple countries' exchanges. In most papers stocks with share prices, of less than 5$ or 3$ were excluded, primarily to avoid micro-structure effects i.e., illiquidity or market manipulation.

The time intervals on which the research was conducted are evenly diversified. For the ranges of a period in years: (0; 1], (1; 5], (5; 10], (10; 20], (20; 40], (40; 100), the number of articles is between 9 and 15. Additionally, four studies analyzed data divided into subperiods, i.e., the whole sample period was divided, for example, into the pre-crisis period and the post-crisis period. Data frequency was mostly monthly (30 studies) and daily (33 articles). Besides, four research were conducted on weekly data, and one on quarterly data or comparing results for daily, weekly, and monthly intervals. Two studies were conducted on intraday intervals, i.e., on half-hour and hour data.

The investor sentiment measures used in the research differed significantly among studies, but one may notice some similarities. The BW index appeared in 12 articles as the only measure of sentiment in a paper. Only measures based on sentiment from the media, i.e., newspapers, internet forums, reports from companies' statements, etc., were more frequent than the BW index, namely appeared in 17 research. However, there is no consensus regarding the method used to develop such indicators. The Naïve Bayes classification appeared several times, although in most cases the researchers used various dictionaries or their own created algorithms. Sentiment measures based on data from Google or Twitter occurred with a similar frequency, i.e., five and four, respectively. Important from the point of view of this review is the fact that 14 articles contained at least two measures, often including the BW index and indirect measures. The rest of the analyzed articles, i.e., 19, focused on various measures such as VIX, closed-end fund discount, or survey proxies. In most cases, these were single measures of sentiment.

---

[2] Please note that the sum of the number of articles and percentages shares when divided by models applied and asset used may exceed 71 and 100%, respectively, because most of the studies cover at least two of these assets.

The diversity can also be seen in the models used in studies. Most often, in 27 papers, the researchers used several different multifactor models, i.e., CAPM, FF three-factor model, and the Carhart four-factor model. In such comparison articles, there were also very elaborate models, i.e., five- and six-factor models. Only 13 articles used only one model of the commonly known ones, i.e., CAPM, FF three-factor model, or Carhart four-factor model. In 22 cases, medium complex models were used, i.e., the model that considers sentiment and additional control variables that do not appear in known multifactor models, such as volume, day of the week, or macroeconomic variables. 19 studies applied only single-factor models, i.e., the model that considers only a sentiment. In recent years also, ML models (7 papers) have been applied for including investor sentiment in asset pricing.

**Table 1:** The distribution of analyzed articles by various features.

| Characteristic | Number of articles | Percentage share of articles in this study |
|---|---|---|
| **Geographical scope** | | |
| The U.S. | 54 | 76% |
| Multiple countries | 10 | 14% |
| Another single country | 7 | 10% |
| **Time interval (in years)** | | |
| (0; 1] | 9 | 13% |
| (1; 5] | 13 | 18% |
| (5; 10] | 12 | 17% |
| (10; 20] | 11 | 15% |
| (20; 40] | 15 | 21% |
| (40; 100) | 11 | 15% |
| **Data frequency** | | |
| Intraday | 2 | 3% |
| Daily | 34 | 48% |
| Weekly | 4 | 6% |
| Monthly | 30 | 42% |
| Quarterly | 1 | 1% |
| **Sentiment measure** | | |
| Media-based sentiment | 17 | 24% |
| BW index | 12 | 17% |
| Multiple measures | 11 | 15% |
| Google Search Volume | 5 | 7% |
| Twitter | 4 | 6% |
| Other measures | 14 | 20% |
| **Models** | | |
| Single factor | 18 | 25% |
| Medium complex | 20 | 28% |
| Multifactor | 30 | 42% |
| Machine learning | 8 | 11% |
| **Asset** | | |
| NYSE | 14 | 20% |
| NASDAQ | 12 | 17% |
| Various countries' indices | 11 | 15% |
| S&P500 | 9 | 13% |
| All stocks from CRSP | 8 | 11% |
| IPOs | 6 | 8% |
| DJIA | 5 | 7% |
| Other assets | 19 | 27% |

*Source: The data in the table has been prepared based on articles specified in detail in the bibliography.*

Due to the extensive scope of the subject and the multiplicity of approaches in the analyzed studies, two analyzes were carried out. The first, qualitative, is based on a condensed description of the results obtained, drawing consistent conclusions from research, presenting unverified gaps, comparing measures, etc. The second, quantitative, relies on an attempt to quantify the overall relationship between investor sentiment and stocks' returns and the improvement of the accuracy of models (mainly using adjusted $R^2$ and frequencies).

*4.1 Qualitative analysis*

The analysis presented in this section comprises a description of each group of models. The fundamental division is based on the complexity of the models. The first group includes single-factor models. The second group includes medium-complex models that use various other factors, excluding those present in multifactor models, such as macroeconomic variables, in addition to sentiment. The third group describes multifactor models, such as the FF three-factor model or the Carhart four-factor model. The fourth group presents models from articles that compare at least two sentiment indicators. It is noteworthy that the papers described in this subsection were not previously mentioned in the previous three subsections, despite the possibility that the models employed could have been assigned to one of the earlier subsections. This is because the aim is to highlight whether comparative studies can establish the superiority of one measure over another without considering these studies in two chapters. Next, the ML models are introduced, followed by models that analyze data from IPOs.

*4.1.1 Single-factor models*

Table 2 presents the summary of sentiment measures used in single-factor models. In such research, mostly media-based and unusual measures were examined. Only one study, i.e., Fisher and Statman (2003) regressed NASDAQ and S&P 500 returns on known direct measures of sentiment, i.e. customer confidence measures of the MSCI index and Conference Board. There were statistically significant relationships between some components (e.g., expectations) of customer confidence and subsequent NASDAQ and small-capitalization stock returns. However, this relationship was not significant for S&P 500 returns. For media-based measures results were significant except for two measures. Seven out of eleven studies showed positive sentiment impact, three presented negative impact, and one revealed both depending on the measure applied. Moreover, in some studies, return reversals have also been observed in a short period for media-based measures. Unusual measures proved to be significantly negative, and one study revealed an insignificant effect. Generally, the single-factor models showed that it is more common to meet and demonstrate significance in unusual measures such as results of a soccer game than in direct ones (in that case the MSCI index and Conference Board measure).

**Table 2:** Summary of various sentiments, i.e., their characteristics, frequency, and the collective results obtained in studies regarding single-factor models.

| Sentiment measure | Description | Frequency | Results |
|---|---|---|---|
| **Media-based measures** | A sentiment measure based on textual analysis (e.g., using Bayes classifier) from various sources, i.e., forums (Yahoo's message board), news services, microblogging platforms, or social media. | Intraday – 1 Daily – 4 Various – 1 Sum – 6 | The results were mostly significant (even on intraday intervals) except for the semantic sentiment based on Yahoo's message board and SA comments. Some studies confirmed the reversal effect of sentiment. |
| **Other** | Customer confidence measures of the MSCI index and Conference Board, Facebook's Gross National Happiness Index, soccer game results, or disease spread. | Daily – 1 Monthly - 4 Sum – 5 | Most research showed a significant coefficient for sentiment except for one for S&P 500 returns. Most often the negative aspect of sentiment was captured, while the positive one (e.g., winning a match) was insignificant. |

*Source: The data in the table has been prepared based on articles specified in detail in the bibliography.*

Very often a linguistic sentiment was studied in single-factor models. Das and Chen (2007) analyzed tech-sector stocks consisting of the Morgan Stanley High-Tech Index. They regressed the value of the index on its first lag and the first lag created by them semantic sentiment on Yahoo's message board. The regression showed that the index and individual stocks were weakly related to the sentiment from the previous day at 10% significance and not significant at all, respectively. Thus, these results implied that aggregation of individual stock sentiment may result in a reduction of idiosyncratic error in sentiment measurement.

However, further studies showed different results. Kim and Kim (2014) studied a sentiment based on the Yahoo! Finance message board using the Naïve Bayes classification (following the Antweiler and Frank (2004) algorithm). They found positive coefficients for all sentiment variables used in any horizon, irrespective of controlling for size and BM ratio. Chen et al. (2014) performed an analysis of daily returns for sentiment indicators based on frequencies of words from Dow Jones News Services (DJNS) and Seeking Alpha (SA) articles and comments. Measures coming from the SA were significantly negative at the level of 5% and the DJNS indicator was insignificant. The predictability is held even after controlling for the effect of traditional advice sources such as financial analysts and news media predictions. The results were different compared to the literature (Tetlock 2007; Tetlock et al., 2008) due to the difference in the return window. DJNS articles are news articles and, as such, can be expected to have more of an immediate impact on prices. SA articles and comments, on the other hand, reflect more of a medium- or long-term view. Garcia (2013) conducted the longest study regarding a sentiment analysis, i.e., over the period 1905-2005 on the DJIA index. As a sentiment indicator, the fraction of positive, negative, and pessimistic (i.e., the difference) words in two columns of financial news from the New York Times was used. Only the first lag for negative, positive, and pessimistic proxies were significant. For positive and negative news, the effect reverses after 5 days, i.e., the sum of betas was statically not different from zero, while for pessimism this sum was positive.

Sometimes researchers also applied data from social media. Sprenger et al. (2014) analyzed the impact of stock-related messages from Twitter on the S&P100. The regression in returns showed a significant positive relationship with bullishness and agreement among tweets. However, they did not find message volume to be related to stock returns. Siganos et al. (2014) examined the relationship between sentiment and stock returns within 20 international markets using Facebook's Gross National Happiness Index. The research showed a significantly positive impact of sentiment on stock returns for small-large, premium, value, and growth. However, after including up to 5 lags for return and sentiment a coefficient for a sentiment become less significant, being even insignificant for Europe countries and marginally (at 10%) significant for American ones.

One study tested the relationship between SP500, DJIA, and NASDAQ returns and investor sentiment from messages posted on the microblogging platform StockTwits in intraday intervals (Renault, 2017). With control of the past market returns, it was found that the first half-hour change in an investor sentiment positively and significantly predicts the last half-hour S&P 500 return. This finding provided evidence that the intraday sentiment effect is distinct from the intraday momentum effect.

Some researchers apply more unconventional sentiment indicators. Edmans et al. (2007) using indices for 39 countries tested the relationship between its returns and international soccer results. They found that: 1) national stock markets earn a statistically significant negative return on the day after a loss by the national soccer team, 2) the loss effect is stronger in small-capitalization indices, and 3) the loss effect is of the same magnitude in value and growth indices. The research did not show any effect of winning. Similar results were achieved after examining a relationship between game results and returns on twenty UK soccer clubs listed on the LSE. A positive relationship between goal difference and win in all cases, i.e., in abnormal results and cumulative abnormal results up to three days, and negative for loss only for cumulative abnormal returns for 3 days ahead were found (Palomino et al., 2009). Even health data can act as a proxy. Liu et al. (2020) tested the effect of daily COVID cases on 21 stock market indices in major affected countries. They found a significant negative confirmed COVID case and that countries in Asia experienced more negative abnormal returns compared to other countries.

*4.1.2 Medium complex models*

Table 3 describes the results obtained for sentiment measures used in medium complex models. Medium complex models applied more diverse measures in comparison to those used in single-factor models. All the results were significant with some exceptions for cases e.g. in bearish market conditions. In this type of model, eight of them showed a negative effect of sentiment, 6 presented a positive impact, and one revealed both depending on the period tested. Also, the reversal effect was observed, the persistence of the effect regardless of the division into subsamples, conditioning the occurrence of size and momentum effects, as well as the moderation of the results by bearish/bullish market conditions.

**Table 3:** Summary of various sentiments, i.e. their characteristics, frequency, and the collective results obtained in studies regarding single-factor models.

| Sentiment measure | Description | Frequency | Results |
|---|---|---|---|
| **BW index** | The first principal component of the following six sentiment proxies suggested by prior research: the closed-end fund discount, market turnover, number of IPOs, average first-day return on IPOs, equity share of new issuances, and the log difference in BM ratios between dividend payers and dividend non-payers. | Daily – 1 Monthly – 1 Sum – 2 | The BW index sentiment conditioned the occurrence of size and momentum effects. The sentiment had no statistically significant effect in the bearish market environment, while it had a negative impact in the bullish conditions. |
| **EPU** | Economic policy uncertainty (EPU) is a risk in which policies and regulatory frameworks are uncertain for the near future. | Monthly – 2 | The EPU had a significant negative effect on stock returns in both studies. One also showed that sentiment was significant for the full sample, before and after the crisis for almost all quantiles. |
| **Direct measures** | AAII's survey shows the percentage of investors who are bearish, bullish, or neutral on stocks. Consumer Confidence reflects consumer attitudes, buying intentions, and consumer expectations for stock prices, inflation, and interest rates. | Daily – 1 Monthly – 1 Sum – 2 | The first study found that both measures were significantly negative for a bear market. The second showed a negative relationship for value, growth, and small stocks for different forecasting horizons. There were insignificant results for large stocks in all horizons and size premiums for 12M and 24M horizons. |
| **Google SVI / FEARS** | Google SVI shows how often a specific term is searched about the total search volume globally, within a defined date range. The Financial and Economic Attitudes Revealed by Search (FEARS) index aggregates daily search volume for keywords related to a household's economic and financial situation. | Daily – 1 Weekly – 1 Monthly – 1 Sum – 3 | Studies revealed a significant coefficient on the SVI for the first and the second weeks. The relationship was not present for further weeks ahead. One study showed that returns from weeks 5 to 52 were negatively related to SVI. Another included the subprime crisis the period after the crisis. The reversal effect was only present in the first period. FEARS were negatively significant when SVI was also present in the regression. |
| **Other** | One media-based index, two measures related to weather, one per aviation disaster, Ramadan, and the spread of disease. | Daily – 4 Monthly – 1 Sum – 5 | For 39 countries a news variable was insignificant, while its interaction with returns was significant, but after including lags of returns. In one of two studies using weather variables indicator was significant. Ramadan and the spread of disease were significant. |

*Source: The data in the table has been prepared based on articles specified in detail in the bibliography.*

Starting with the simplest measure, Chen et al. (1993) tested the monthly changes in the discount of a value-weighted portfolio of closed-end fund discount on NYSE stock returns for three different periods, i.e., 1965-1975, 1975-1985, and the whole period. For the first and the third, the coefficient was negative and significant, whereas for the second it was insignificant. While Kurov (2010) regressed the BW index on excess returns on stocks from the S&P500 in two market conditions, i.e. under a regime with a higher mean and lower variance of returns (bull market) and a regime with a lower mean and higher variance (bear market). In the bear market environment, a sentiment had no statistically significant effect, while in a bull market condition was significantly negative. The study also analyzed the term spread factor, which was significant in the same way as the sentiment. This finding is consistent with monetary shocks having little effect on stocks in good times. These findings support the conclusion that Fed policy affects stock returns, at least in part, through its effects on investor sentiment and expectations of credit market conditions. Moreover, Antoniou et al. (2016) using NYSE, AMEX, and NASDAQ data observed that the BW index sentiment conditions the occurrence of size and momentum effects. The size effect was present only in pessimistic periods, while the momentum effect is present only in optimistic periods. While the value effect is present in both periods.

There were also similar studies measuring the influence of uncertainty on stock returns. You et al. (2017) analyzed stock prices from the Resset Financial Database China's industry data before and after the subprime crisis using quantile regression. As a sentiment indicator, they use the EPU index from Baker et al. (2014). They found significant negative effects on stock returns for the full sample, before and after the crisis for almost all quantiles. The impact on stocks in pre-crisis was relatively greater than that in post-crisis at most quantiles. While Chen et al. (2017) analyzed the relationship between all A-share stocks listed in Shanghai and Shenzhen stock exchanges and China's EPU. For regression without any controls and including economic and market uncertainty variables, authors found EPU's significant and negative relationship. An out-of-sample predictions study showed that the MSFE-t and MSFE-F statistics were both statistically significant at the 10% level at least.

In two studies researchers used direct proxies for a sentiment. Kurov (2008) examined AAII and II's sentiment index on the S&P 500 and Nasdaq-100. He regressed returns for the bull and the bear market using the beta factor for the market, the default spread factor, the term spread factor, and sector dummies. He found that both measures were significantly negative for a bear market. Schmeling (2009) examined the relationship between consumer confidence (as a proxy for investor sentiment) and 18 developed countries' stock returns. The sentiment negatively forecasts aggregated stock market returns on average across countries (for 9 countries on a 5% level of significance and 11 countries on a 10%-level only). The higher the sentiment, the lower the future stock returns and vice versa. This relation is held for value, growth, and small stocks for different forecasting horizons. However, there were insignificant results for large stocks in all horizons and size premiums for 12M and 24M horizons.

A couple of studies examined the Google SVI index. Da et al. (2011) found a significant positive coefficient of the SVI on Russel 3000 for the 1st and the 2nd weeks after controlling for alternative measures of investor attention. The relationship was not present for the 3rd, 4th, or between the 5th and 52nd weeks ahead. Also for the first two weeks, the study showed negative interaction between equity market capitalization and a positive with a retail trading volume measure for the first week. However, none of these relationships were present for the 3rd, 4th, and between the 5th and 52nd weeks ahead. The authors additionally examined long-run returns, i.e. following Barber and Odean (2008), they skip the 1$^{st}$ month and look at the returns from weeks 5 to 52 and found a negative coefficient of the SVI, like the magnitude of total initial price pressure in the first two weeks, suggesting that the initial price pressure was almost entirely reversed in 1 year. However, the negative coefficient is marginally insignificant. The same researchers in 2015 performed regressions on S&P500 return every six months also on Google SVI, but this time creating a Financial and Economic Attitudes Revealed by Search (FEARS) index. Such a measure was negatively and significantly related to returns when both were observed at the same time. However, FEARS occurred to be significant and positive for returns in t+1, t+2, and from t+1 to t+2 periods. That supports the reversal nature of the sentiment, while the cumulative impact of an increase in FEARS predicts a cumulative increase of returns over days 1 and 2. This was significant after controlling the EPU, VIX, and changes in the Aruoba-Diebold-Scotti business conditions index. Although, the model does not predict returns for times t+3, t+4, and t+5. Bijl et al. (2016) examined the effect of Google SVI on the S&P 500 in two periods. One included the subprime crisis (2008-2013), whereas the letter covered the only period after the crisis (2010-2013). For the first period, the authors found a significant negative coefficient for search volume for its first and third lags, and a significant positive coefficient for the second lag. While the second-period coefficient was significantly negative only for the first and fourth lag. Then the reversal of prices was observed only in the full sample period.

In medium complex models also the media-based indicator was used. Klibanoff et al. (1998) on a sample of country funds consisting of 39 single-country publicly traded funds applied major news events using the column width of front-page articles in the New York Times. The regression was conducted on the fund's returns one week ahead. They found that a news variable was never significant (in any proposed model).

Studies also applied unconventional proxies for a sentiment. Goetzmann et al. (2015) examined the impact of the sky cloud cover on stocks from CRSP in subsamples based on arbitrage costs. The study showed that more cloudy days increased perceived overpricing in individual stocks and the DJIA Index, and selling propensities of institutions. Based on that the authors introduced stock-level measures of investor mood. The investor optimism positively impacted stock returns among stocks with higher arbitrage costs. These findings complement existing studies on how weather impacts stock index returns and identify another channel through which it can manifest. Similarly, Chang et al. (2008) examined NYSE stocks and the weather in New York City, i.e. wind speed, snowiness, rainfall, and temperature. In general, weather variables are not significantly related to returns. Kaplanski and Levy (2010) examined the effect of aviation disasters on NYSE stock prices. When regressed on the lagged rate of returns and other controls disasters event coefficient on the first day was significantly negative significant, while on the second day was insignificant. The effect was greater in small and riskier stocks and firms belonging to less stable industries. Białkowski et al. (2012) analyzed the impact of Ramadan in 14 Muslim countries. The positive and significant effect of Ramadan materialized only when the society chooses to participate in this religious experience collectively, i.e. at least 50% of citizens were Muslims. Ichev and Marinč (2018) examined whether the geographic proximity of information disseminated by Ebola outbreak events with intense media coverage affected stock prices of NYSE and NASDAQ indices. The negative event effect was the strongest for the stocks with exposure of their operations to the African countries and the U.S. Moreover, the events located in these regions were also the strongest. This result suggests that the information about Ebola outbreak events is more relevant due to the geographical distance to both the place of the Ebola event and the financial markets. The effect was greater for small and more volatile stocks, stocks of a specific industry, and stocks exposed to intense media coverage.

*4.1.3 Multifactor models*

The results obtained for sentiment measures used in multifactor models are briefly described in Table 4. The most popular measure applied in such models was the BW index. For this measure, all studies, except for one, showed significant results in explaining various returns. The second frequently observed a group of measures were media-based ones. Like the BW index only one study showed insignificant results. Other measures of sentiments also revealed significant results, however, they were not studied in more than one research. Seven research revealed a negative effect of sentiment, nine showed a positive impact, and the rest of the studies (8) presented both signs depending on the decile or anomaly/strategy tested. To sum up, for almost every multifactor model, investor sentiment turned out to be important, which emphasizes its importance and indicates that its impact cannot be explained by fundamental factors. The models described in this subsection specifically indicate that augmenting models with the investor sentiment proxies improves the accuracy of models. Thus, they support the first hypothesis.

**Table 4:** Summary of various sentiments, i.e. their characteristics, frequency, and the collective results obtained in studies regarding single-factor models.

| Sentiment measure | Description | Frequency | Results |
|---|---|---|---|
| **BW index** | Explained in Table 3. | Daily – 2<br>Monthly – 8<br>Sum – 10 | The BW index explained portfolios' returns based on BM ratio, size, dividend premium, volatility, R&D expense, sales growth, MAX factor (see Bali et al., 2011), profitability, and external finance. The indicator was also important in explaining most of the anomalies, in particular on the short side of the portfolios. One study showed that using the Carhart model the index was insignificant in regression on momentum returns. |
| **Media-based measures** | Explained in Table 2. | Daily – 3<br>Monthly – 3<br>Sum – 10 | All measures (both based on tone and frequencies) applied were significant explaining excess returns and momentum returns. One study showed that media coverage was insignificant in regression on losers and mid returns, but only for the winners. |
| **Other** | The CEFD, the opening accounts number and turnover rate, terrorism events, the VIX, the buy-sell imbalance, the EPU, and Google SVI. | Intraday – 2<br>Daily – 2<br>Weekly – 2<br>Monthly – 3<br>Sum – 11 | The CEFD was significant in regression on returns divided in deciles by equity value. The opening accounts number and turnover rate were significant in the quantile regression conducted on the Chinese stock market. Terrorism events had a significant impact on stock market returns in 22 countries. The VIX was significant in regression |

on portfolios sorted on BM equity, size, and beta. The buy-sell imbalance explained returns for stocks with high retail concentration. The Google SVI was significant in regression on volatility-sorted portfolio deciles. Some studies confirmed the reversal effect of the sentiment.

*Source: The data in the table has been prepared based on articles specified in detail in the bibliography.*

Some research used the CAPM model or CAPM model with one or a couple of additional dependent variables. Phan et al. (2018) using data from 16 countries tested whether the EPU measures (i.e. country-specific and global one) can predict stocks' returns. They found the predictability of excess returns for 5 countries, where both the country and the global EPU models outperformed the constant model. No predictability was found for 10 countries for the local EPU and global EPU. Lee et al. (1991) tested the effect of the monthly CEFD on NYSE divided in deciles by equity value. The largest firms did significantly poorer when discounts narrowed, while for the other nine portfolios, stocks did significantly better when discounts shrank. When an equal-weighted market index was used, however, the five portfolios of the largest firms all showed negative movement with the value-weighted discount, while the five smaller portfolios all had positive coefficients.

One of the most popular models applied was the FF three-factor model. Tetlock (2007) analyzed the effect of the pessimism media factor from the WSJ column on DJIA using that model. It occurred that the first and the fourth lags of the pessimism factor were negative and significant. Stambaugh et al. (2015) applied the lagged BW index for NYSE, AMEX, and NASDAQ stocks containing stocks with either the highest (top 20%) or the lowest (bottom 20%) idiosyncratic volatility[3]. They found significant negative loadings for the BW index for the highest three quintiles in the highest minus the lowest quintiles and the lowest quintile and all stocks. Chung et al. (2012) regressed the BW index on the same portfolio's returns. For single and multifactor regressions, the study showed similar results, i.e. positive loading for the long-short portfolios based on size, BM value, age, earnings, and dividend premium, and negative for volatility, R&D expense, sales growth, and external finance. Corredor et al. (2013) referred all stocks listed in four of the key European markets, i.e. France, Germany, Spain, and the UK to the BW index and EU sentiment measures. The second was constructed from the first principal component of the first factors obtained for each country and then the PCA was used to create an aggregate index. Regressions of long-short portfolios for a 6, 12, and 24-month time horizon were constructed for BM ratio, size, volatility, and dividend premium. The BW index was significant and had the expected size for most of the portfolio, whereas European sentiment mostly was insignificant. Takeda and Wakao (2014) tested the impact of Google SVI on the Nikkei 225 index. They found that the coefficient on the search intensity was significantly positive. Jacobs (2015) found that the BW index is a powerful predictor for most anomaly returns (out of 100), on the short side of the portfolios. Ni et al. (2015) used an opening accounts number and a turnover rate to constitute the investor sentiment. They employed the quantile regression model to verify the effect of investor sentiment on monthly stock returns in the Chinese stock market. The findings showed that the influence of investor sentiment was significant from 1 month to 24 months. The effect was asymmetric and have a reversal nature, i.e. it was positive and large for stocks with high returns in the short term and negative and small in the long term. Drakos (2010) explored whether terrorism events have a significant negative impact on daily stock market returns in a sample of 22 countries. The terrorist activity had a negative impact and reduced significantly daily returns even after controlling for global financial crises.

---

[3] An idiosyncratic volatility measures the part of the variation in returns that cannot be explained by the particular asset-pricing model used.

Some studies considering the FF three-factor model showed insignificant results. Hribar and McInnis (2012) used the BW index as a dummy variable equal to 1 if the beginning of the year sentiment index was positive, and 0 in other cases. Findings showed that such an indicator is significant in predicting young minus old, volatile minus smooth, and nonpayers minus payers' stock returns. However, after including the FF three-factor in the regression proxy became insignificant. Han et al. (2013) verified the BW index on NYSE and AMEX stocks' returns. The findings showed that the coefficient for the sentiment index in the FF three-factor model was insignificant.

Even more often than the FF three-factor model researchers employed the Carhart four-factor model. Baker and Wurgler (2006) studied how their newly created investor sentiment index affects the cross-section of stock returns. They created long-short portfolios based on low, medium, and high firm characteristics, where low is defined as a firm in the bottom three NYSE deciles, high in the top three NYSE deciles, and medium in the middle four NYSE deciles. The study showed that when sentiment at the start of the period is low, subsequent returns are relatively high for stocks with low market capitalization, low age, high volatility (i.e. the annual standard deviation in monthly returns for the last 12 months), unprofitable (i.e. with net income lower than zero), and dividend-free. For the growth and distress variables (i.e. external finance over assets and sales growth) the results did not show simple monotonic relationships with a sentiment. For both low and high sales growth and external finance over an asset, returns are low relative to returns on a medium of these characteristics. Whereas, when sentiment is high, these stocks earn low. They found that the size effect of Banz (1981) appears only in low sentiment periods. The sentiment was negative for size and volatility long-short strategies and positive for age long-short strategies. Fong and Toh (2014) examined the BW index on NYSE, AMEX, and NASDAQ returns. They regressed excess returns of the long-short MAX (see Bali et al., 2011) portfolio against the lagged BW sentiment index for each institutional ownership (IO) quintile controlling Carhart's four factors plus liquidity risk factor. Returns on the portfolios were negatively related to the sentiment proxy for most IO quintiles except for the third and the fifth quantiles. Mian and Sankaraguruswamy (2012) used the BW index on CRSP stock returns. They performed regression on sentiment and Carhart's four factors. The sentiment was negatively related to the difference in returns between the high and low news stocks. However, Moskowitz et al. (2012) using quarterly data for nine equity indexes from developed markets showed on the Carhart model that the BW index of sentiment and its extreme values (top 20% / bottom 20%) were insignificant in regression of time series momentum returns on the market.

Researchers used also textual measures in the Carhart four-factor model. Bartov et al. (2018) investigated the relationship between the aggregate opinion in individual tweets and Russell 3000 index using the same model. They found a significant and positive relationship between Twitter opinion and returns around earnings announcements with various controls. After controlling the factors the effect persisted only for volatility and age-based portfolios. Joseph et al. (2011) examined the ability of online ticker searches in the SVI to forecast S&P 500 abnormal stock returns on volatility-sorted portfolio deciles. The betas associated with the sentiment indicator generally increased as the volatility grew, starting from a negative value at the first decile and finishing at a positive value for the tenth decile. The letter was greater in absolute value as compared to this from the first decile. Xiong and Bharadwaj (2013) obtained the firms' monthly frequencies of news data from Lydia/TextMap (Lloyd et al., 2005). They regressed those frequencies on abnormal returns got from the Carhart four-factor model. They observed that positive and negative news had significant effects on returns. The interaction between positive news and advertising was positive, while for negative news this interaction was insignificant. Yu et al. (2013) used a web crawler to download blogs, forums, and news web pages and applied the Naïve Bayes algorithm to conduct sentiment analysis. They got abnormal returns from the model and run fixed effect regression on volumes with interactions. Findings suggested that generally social media had a stronger relationship with stock returns than conventional media. Whereas social and conventional media had a strong interaction effect on stock performance. Moreover, the impact of different types of social media varied significantly.

Some studies analyzed other measures than the BW index or textual data in the model. Bartov Banerjee et al. (2007) wanted to find whether the VIX predicts returns on stock market indices (NYSE, AMEX, and NASDAQ). They examined portfolios sorted on BM equity, size, and beta with controlling of the four Carhart four factors. The coefficients were positive and significant except for portfolios based on the low beta, the low BM value, and the large size. Kumar and Lee (2006) using the buy-sell imbalance of more than 1.85 million retail investor transactions over 1991–1996 showed that systematic retail trading can explain return comovements for stocks with high retail concentration (i.e. small capitalization, value, lower institutional ownership, and lower-priced stocks), especially if these stocks are also costly to arbitrage.

There was research comparing different multifactor models, as well. Fang and Peress (2009) examined the relationship between the number of newspaper articles about a stock (coming from the LexisNexis database) with NYSE and NASDAQ stocks. The difference between the no- and high-coverage groups is statistically significant and economically meaningful. In the regressions on long no-media stocks and short high-media stocks CAPM, FF 3-factor, and Carhart 4-factor all factors were significant. Hillert et al. (2014) tested whether stocks traded on NYSE, AMEX or NASDAQ can be related to firm-specific articles from newspapers from the LexisNexis database. They calculated media coverage as a frequency, and tone came from a textual analysis following the dictionary approach developed by Loughran and McDonald (2011). They computed different risk-adjusted (i.e. CAPM, 3F, 4F, 6F) momentum returns for stock portfolios sorted by residual media coverage based on a holding period of six months. They showed that firms covered by the media exhibited stronger momentum depending on the tone. That effect reversed in the long run and was more pronounced for stocks with high uncertainty characteristics. These results collectively lent credibility to an overreaction-based explanation for the momentum. However, media coverage did not change losers and mid-returns, but only for the winner. Stambaugh and Yuan (2017) regressed for five-factor alternative models on NYSE, AMEX, and NASDAQ excess returns on either the long, short, or long-short leg for the following factors: market, SMB, MGMT and PERF, and the BW index. For MGMT[4] and PERF[5], the coefficients on short legs are uniformly negative and positive for long-short. The slopes for market and SMB were insignificant.

### *4.1.4 Multiple indicators*

Some research considered comparing individual proxies' performance in asset pricing models. Neal and Whitley (1998) using extensive data from 1933 to 1993 for NYSE-AMEX analyzed the impact of the closed-end fund discount, the ratio of odd-lot sales to purchases, and the net mutual fund redemption on stocks returns. They found that fund the first one and the last predicted the size premium, but the odd-lot ratio did not. Brown and Cliff (2005) investigated the impact of the II survey results and closed-end fund discount, the ratio of NYSE odd-lot sales, the net mutual fund flows, the ARMS index (a popular measure of sentiment among technical analysts), the number and returns on IPOs on DJIA stock returns. Coefficients are almost universally significant and negative and tend to be most negative for the larger and growth firms. For these firms, sentiment is a significant predictor of future returns at the 1-, 2-, and 3-year horizons. When including all variables together, the survey indicator of sentiment remained significant. There was no evidence that the closed-end fund discount is related to subsequent stock returns. Simon and Wiggins (2001) analyzed the S&P 500 futures contract with indicators including the VIX, the put-call ratio, and the trading index. All the proxies were positive and significant. Lemmon and Portniaguina (2006) explored the time-series relationship between investor sentiment and the small-stock premium using the MSCI index and the Conference Board survey of consumer confidence as a measure of investor optimism. In the period before 1977, the measures were insignificant, however, after 1977 for 3,6,12 months periods there were significant and negative coefficients. The estimate for the interaction between the customer confidence measure and the return on the market index was negative and statistically significant.

Other research compared more complex sentiment indicators such as the BW index. Ben-Rephael et al. (2012) tested the lagged MSCI index, the lagged BW index, and the lagged aggregate net exchanges of equity funds on a value-weighted index composed of NYSE, AMEX, and NASDAQ stocks. The results showed that MCSI and VIX were statistically significant and positive, while the BW index was insignificant. Stambaugh et al. (2012) explored the role of the BW and the MCSI indices on NYSE, AMEX, and NASDAQ in a broad set of anomalies in cross-section stock returns. The measures were significant in most of the anomalies, however, the BW index was more often significant and had a greater value of a t-statistic. Huang et al. (2015) proposed a new investor sentiment proxy created using the Partial Least Squares (PLS) procedure sentiment index from the six individual proxies used to create the BW index and compared it with the BW index, the Naive investor sentiment index, and individual proxies. Regression on returns using only sentiment measures revealed that the BW index was insignificant, the naïve one was marginally statistical significance at the 10% level, while the PLS sentiment was significant and negative at the 1% level. Also return on IPOs and EQTI displayed high power in forecasting the excess market returns. Overall, the PLS index beat all the individual proxies and remained statistically significant when augmenting the model with other economic predictors. Moreover, it exhibited stronger predictive power than other measures. Jiang et al. (2019) examined regressions on stock returns on various portfolios sorted on proxies for limits to arbitrage or speculation. Authors used the following proxies for investor sentiment: the BW index, the PLS investor sentiment index, the MCSI index, the Conference Board Consumer Confidence Index, the FEARS indicator, and the manager sentiment index, which was based on the aggregated textual tone of corporate financial disclosures. All the indicators were significant. But only Huang's investor sentiment remained significant, when in regression also the manager sentiment index was present.

---

[4] The MGMT factor is constructed from a set of six anomaly variables that can be directly influenced by a firm's management (Fang and Taylor, 2021).

[5] The PERF factor is similarly constructed from five anomaly variables that represent a firm's performance (Fang and Taylor, 2021).

Comparing the measures of sentiment often ended with all measures being significant. Although sometimes direct measures turned out to be more significant while single indirect measures did not. In most cases, only studies examining out-of-sample accuracy showed some differences. It turns out that the commonly used BW index is not the best indicator, because even combining the same component variables differently can give more accurate results. Such a fact supports the second hypothesis that more complex sentiments have better predictive power than simpler ones.

*4.1.5 Machine learning*

Through the last decade, researchers started to employ ML techniques to include investor sentiment in asset pricing models. Bollen et al. (2011) used two methods to create a sentiment based on Twitter data for DJIA stock returns. The first was OpinionFinder, which measures positive versus negative mood from text content, and the second was GPOMS which measures 6 different mood dimensions from text content. For the first, no effect on prediction accuracy was found compared to using only historical values. While the second "Calm" created the highest prediction, "Sure" and "Vital" reduced prediction accuracy significantly, while "Happy" significantly decreased average MAPE. Ranco et al. (2015) also used Twitter data to calculate sentiment for 30 stocks from the DJIA index. However, they used Support Vector Machine to compute the proxy. The values of cumulative abnormal returns were significantly positive for ten days after the positive sentiment events. The same holds for negative sentiment events, but the cumulative abnormal returns were twice as large in absolute terms. Oliveira et al. (2017) examined more indices, i.e. the S&P 500, the Russell 2000, the DJIA, and the NASDAQ 100, and constructed a couple of variables based on microblogging data from Twitter – bullish ratio, bearish ratio, bullishness index, variation of ratios and agreement. Then applied different ML models. The study found that Twitter sentiment and posting volume were relevant for the forecasting of returns of the S&P 500 index, portfolios of lower market capitalization, and some industries. Mostly the best predictive results were provided by Support Vector Machine. These results confirm the usefulness of microblogging data for financial expert systems, allowing them to predict stock market behavior and providing a valuable alternative for existing survey measures with advantages (e.g., fast, and cheap creation, daily frequency).

Other researchers used various data for the models. Li et al. (2014) constructed lexical sentiment for the CSI 100 list and applied it to the predictive eMAQT model that captures the hidden connections between the input (textual information, public mood, and current stock prices) and the output (future stock prices). The researchers concluded that: 1) representing news articles with proper nouns could achieve a good directional prediction but attain a poor RMSE; 2) the pessimistic public mood had a significant contribution in predicting stock movements; 3) news articles related to restructuring issues are the most predictable. Weng et al. (2018) employed ML models based on Wikipedia hits, financial news, Google trends, and technical indicators for 20 U.S.-based stocks. MAPE was lower for the simulations with no PCA than with PCA. The boosted regression tree and random forest regression methodologies were the most predictive, while the support vector regression ensemble had the lowest performance. Ding et al. (2015) proposed a deep-learning method for event-driven stock market prediction. Results show that our model can achieve nearly 6% improvements in S&P 500 index prediction and individual stock prediction, respectively, compared to state-of-the-art baseline methods. Nguyen et al. (2015) employed historical prices for the 18 stocks and created sentiment measures for them based on various methods. The aspect-based method occurred to have the best performance. Li et al. (2014) based on the stocks listed in Hong Kong Stock Exchange implemented a generic stock price prediction framework and plugged in six different models. They conducted the textual news articles that are then quantitatively measured and projected onto the sentiment space and evaluated the models' prediction accuracy and empirically compare their performance at different market classification levels. Results showed that at all levels, i.e. at an individual stock, sector, and index, the models with sentiment analysis outperform the bag-of-words model in both the validation set and independent testing set.

The above-described results proved that ML algorithms can be applied to increase the predictive power of the asset pricing model, however, they have a major shortcoming. They are difficult to interpret and no one should forget they also can fail as the traditional models.

*4.1.6 IPOs*

Some researchers also applied an investor sentiment on the returns of IPOs. Cook et al. (2006) got all IPOs from the Securities Data Company's New Issues database. They applied the number of news articles that had mentioned the firm's name in the headline(s) and found a strongly significant positive relationship. Cornelli et al. (2006) used prices from the grey market (the when-issued market that precedes European IPOs) to proxy for small investors' valuations for 486 companies that went public in 12 European countries High grey market prices (indicating overoptimism) were good predictor of first-day prices, while low grey market prices (pessimism) were not. Moreover, the authors found that long-run price reversal only follows high grey market prices. This asymmetry occurred because institutional investors could choose between keeping or reselling them when small investors are overoptimistic. Dorn (2009) investigated the IPOs of the Frankfurt Stock Exchange. They applied two investor sentiment measures, i.e. the logarithm of gross When-Issued purchases, and the logarithm of the gross day plus 1 purchase. In the study the regression of excess returns over Dax 100, Nemax 500, industry, size, BM ratio, internet dummy, and High-tech dummy. Both indicators were negative and significant. Da et al. (2011) regressed IPO first-day returns on pre-IPO week abnormal search volume with and without IPO characteristics. In both cases the sentiment proxy was significant.

The research proved that investor sentiment could be applied to explain returns on IPOs. However, in the studies, the authors used mostly unconventional indicators. Thus, we cannot be sure whether these measures reflect the same phenomenon as the popular measures.

*4.1.7 Summary*

Summarizing, the results obtained in the qualitative analysis showed that sentiment is almost always an important factor in asset pricing models. It was significant in at least one of the tested relationships in 65 out of 71 studies. They were significant regardless of the frequency of the data. However, there were also 6 studies showing that the sentiment was completely insignificant. Such research considered the BW index, weather variables, and measures based on media. They were conducted on various data frequencies and mostly for American markers. The most often studies were conducted in the first decade of the XXI century and the research period was considered 10 years period. Interestingly, the research that considered the BW index were covering a wide period, i.e. from 23 to 44 years, and was applied in the multifactor models. A deeper analysis showed that:

1. the BW index is more often relevant when used at the index level than at the level of individual stocks;
2. the BW index was more likely to be significant when used on older datasets.

The above results showed that this indicator is difficult to generalize to all markets and its importance has been decreasing in recent years, which may indicate that there is a need to identify a new indicator. As a reminder, the BW index was proposed in 2006. At the same time, other research (for various sentiment proxies) divided into sub-periods often turned out to be insignificant for the earlier period, which may indicate that investor sentiment is becoming an increasingly important factor. Moreover, generally, the sentiment often exhibited a reversal effect, i.e. the phenomenon in which the effects of the influence of sentiment are at least partially reversed in subsequent periods. This often resulted in the significance of the first and fourth or fifth lags, and no significance in the second and third lags.

There have been a number of developments in the literature regarding the incorporation of investor sentiment in asset pricing models. One of the most notable is the work of Baker and Wurgler (2006), which showed that sentiment-based investor behavior can have a significant impact on stock prices and the cross-section of expected returns. Additionally, many studies have shown that various sentiment indicators can predict stock market returns. For example, studies have used different measures of sentiment such as survey-based measures, media-based measures, and online text-based measures. These studies have found that sentiment indicators can be used to predict stock returns. Furthermore, more recent studies have looked at the impact of machine learning techniques and natural language processing in the context of sentiment analysis and stock market prediction. These studies have found that using these techniques can lead to improved predictions of stock returns.

The results described in this section are difficult to generalize, because the results on many issues were not consistent, such as the significance of the sentiment split by deciles of other variables or in various time horizons. Nevertheless, due to its significance, its coefficient, and influence on R2, it can be concluded that the obtained results confirm the first research hypothesis (RH1) that augmenting models with the investor sentiment proxies improves the coefficient of determination.

*4.2 Quantitative analysis*

*4.2.1 Single-factor models*

Figure 2, 3, and 4 present the beta coefficients for investor sentiment in the single factor, medium complex, and multifactor models with their standard deviation. The tables present the results of the research, which were also not discussed in detail in the above analysis but were only supplementary. Hence, among others, a greater number of studies for single factor models. The research for single factor models presents the greatest dispersion. That is due to the absence of other factors. A sentiment took over some of the influence of rational factors, and additionally, for such studies, new/experimental measures of sentiment were tested more often than in other studies. In studies of medium complex models, the range of results is the smallest, which can also be explained reasonably. These models are often considered macroeconomic variables that undoubtedly affect investor sentiment. After all, the variability of coefficients in multifactor models falls somewhere in the middle.

**Figure 2:** Comparison of beta coefficients of sentiments with their standard deviation in single factor models (meaning a change in stock return by percentage points).

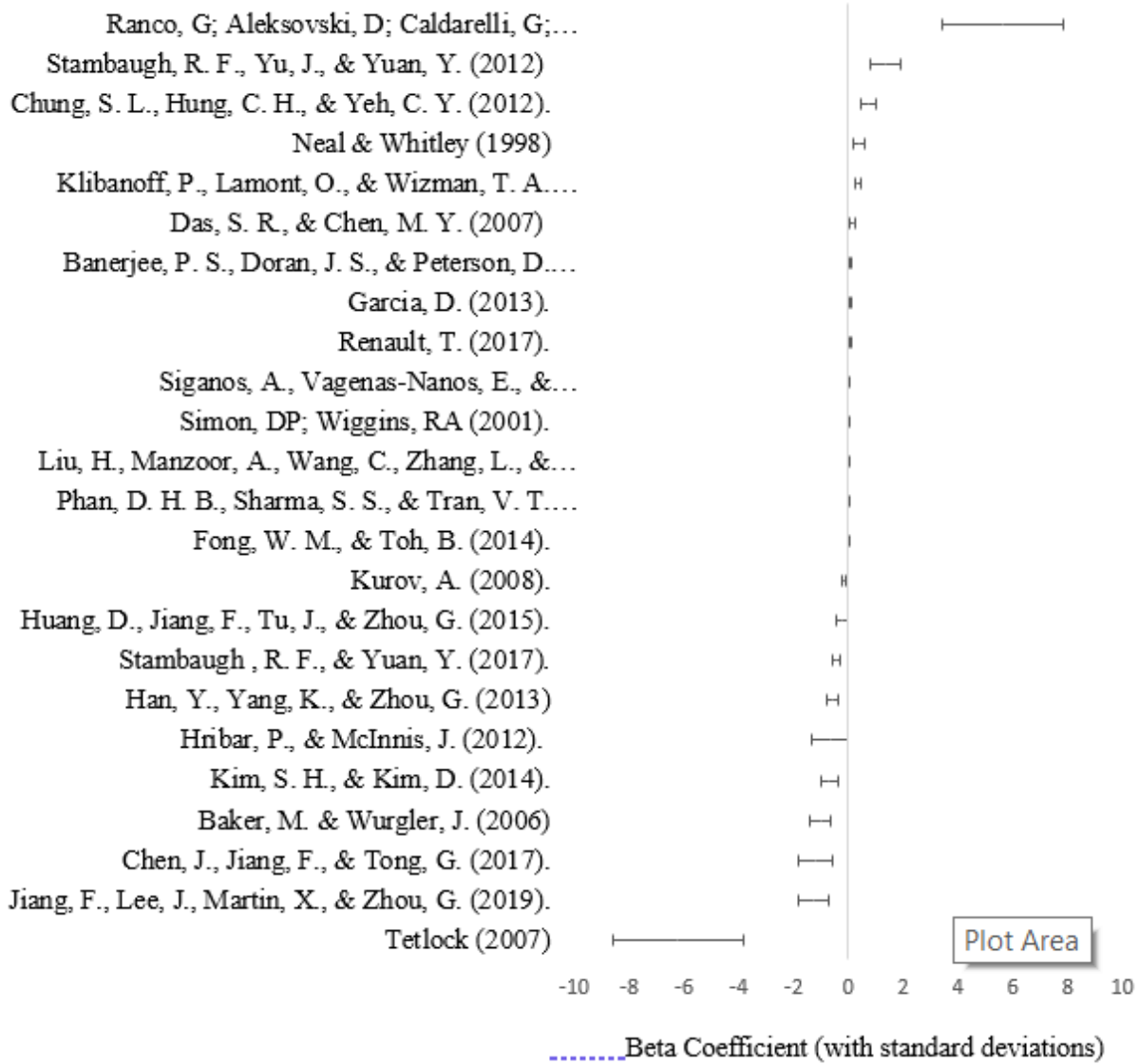

*Source: The data in the table has been prepared based on articles specified in detail in the bibliography.*

**Figure 3:** Comparison of beta coefficients of sentiments with their standard deviation in Medium complex models (meaning a change in stock return by percentage points).

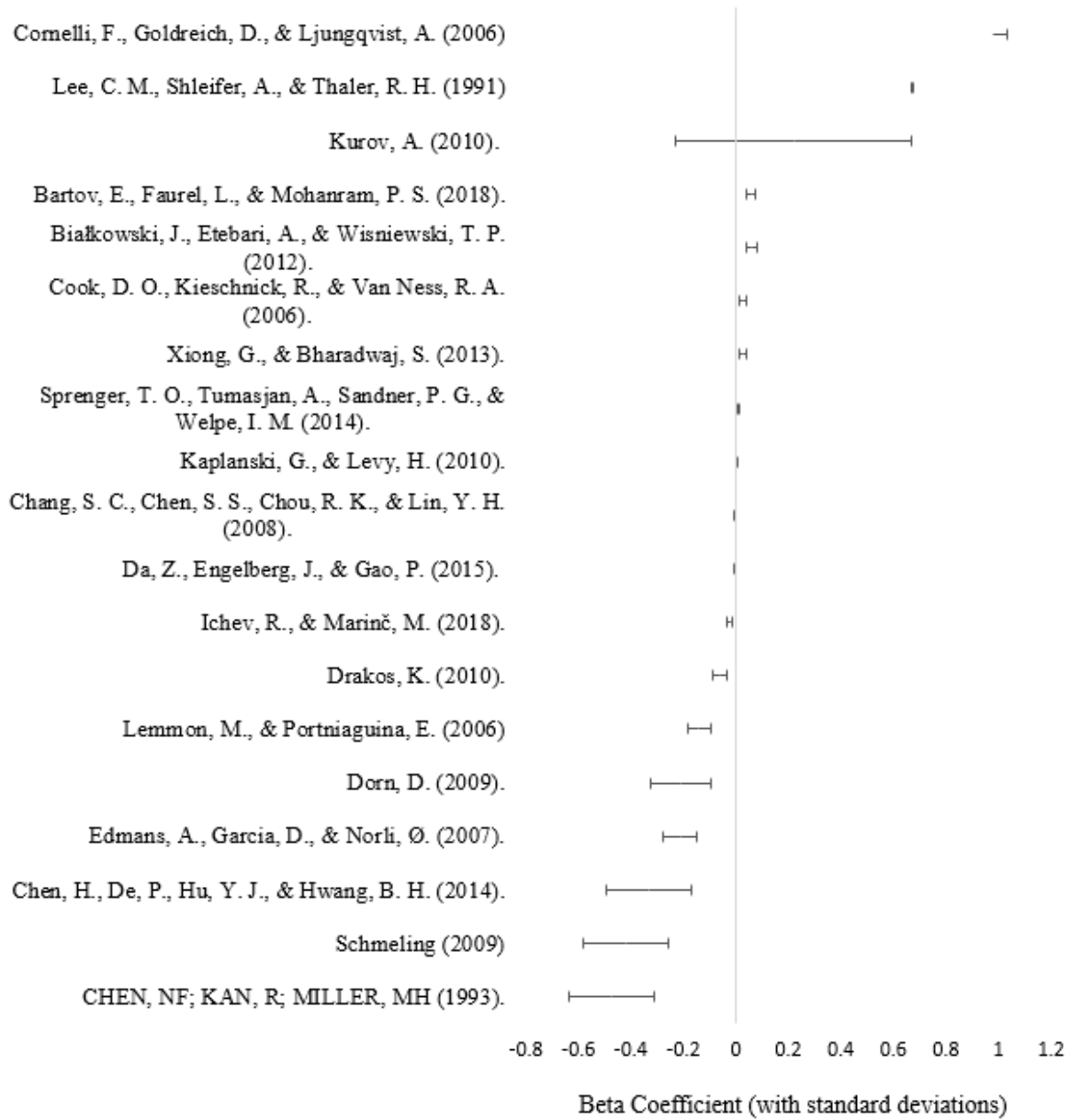

*Source: The data in the table has been prepared based on articles specified in detail in the bibliography.*

**Figure 4:** Comparison of beta coefficients of sentiments with their standard deviation in multifactor models (meaning a change in stock return by percentage points).

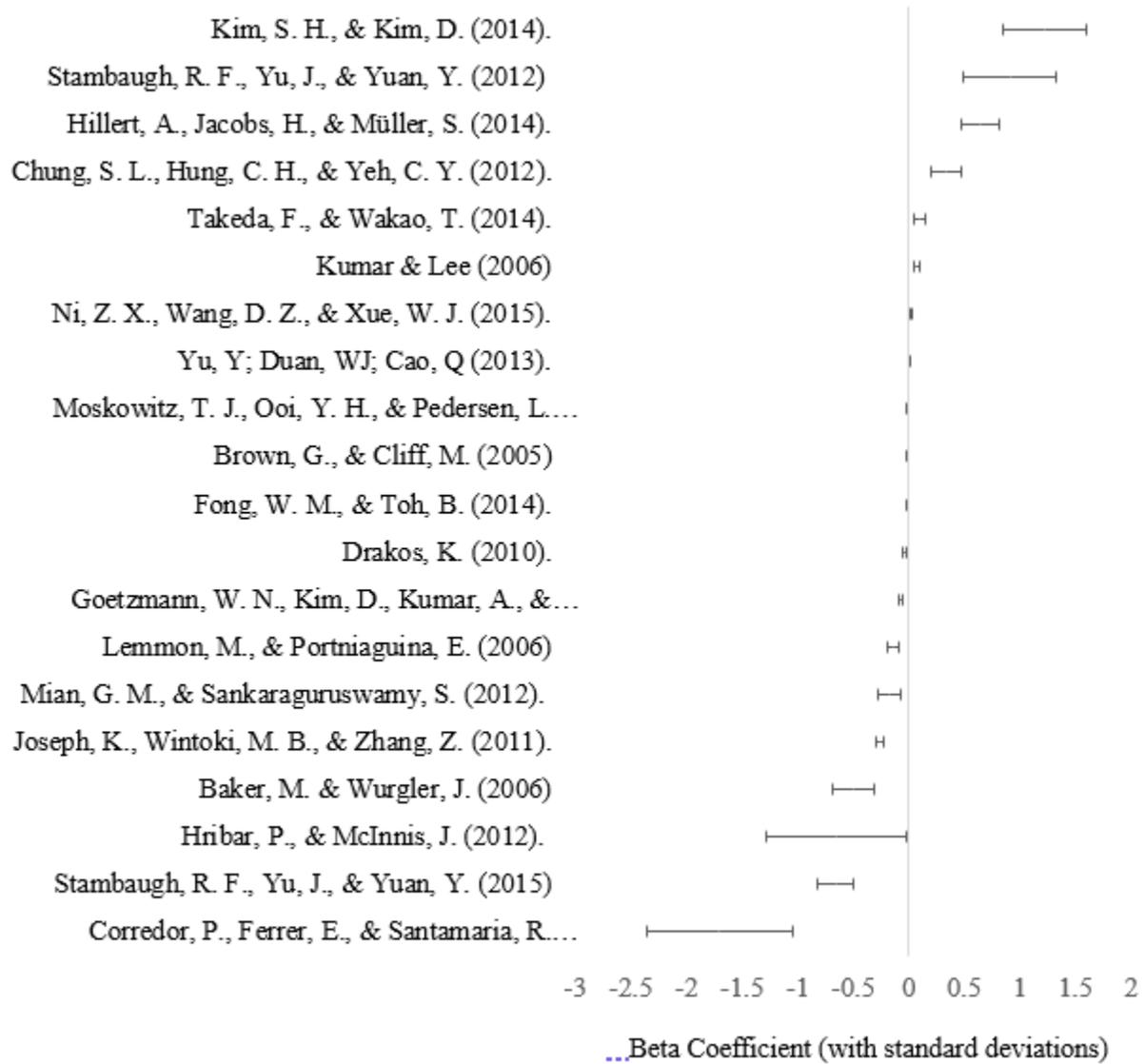

*Source: The data in the table has been prepared based on articles specified in detail in the bibliography.*

Table 5 presents the number of studies in which positive and negative sentiment measures occurred with its coefficients for three groups of models, i.e. single-factor, medium complex, and multifactor models. Note that the number of studies for medium complex and multifactor models is lower by one in comparison to this presented in Table 1 since two studies were only comparing the difference between stocks with high and low media coverage and therefore sentiment was not a factor in regression there.

In general, it was observed that negative sentiment measures were more frequently used than positive ones, although the difference was not significant. It should be noted that for single factor models, the number of studies with sentiment measures having positive coefficients was higher than those with negative coefficients. This suggests that researchers perceive sentiment differently as negative and positive, and they are more interested in studying the reasons behind bearishness in the market. Additionally, the most commonly used sentiment indicator, i.e., the BW index, is a positive proxy, which may explain why researchers are looking for measures that can capture a different phenomenon. However, none of the studies used multiple measures that could capture both the positive and negative effects of sentiment. Moreover, caution must be exercised while interpreting the average coefficient values presented in the table, as they contain both slopes for returns expressed in basis points and raw returns. While single-factor models had a higher number of studies with positive sentiment measures, the average coefficient value was negative, possibly because applied indicators for negative sentiment were more economically significant. It is important to note that no statistical analysis was performed here, given the small number of studies, and the presented conclusions are speculative.

**Table 5:** The number of articles with positive and negative sentiment coefficients divided by the type of model.

| Models | No. of studies with positive sentiment measures | No. of studies with negative sentiment measures | Avg. of coefficients |
|---|---|---|---|
| Single factor | 11 | 7 | -3.02 |
| Medium complex | 6 | 13 | -0.52 |
| Multifactor | 12 | 18 | -0,27 |
| All | 29 | 38 | -1,10 |

*Source: The data in the table has been prepared based on articles specified in detail in the bibliography.*

The first hypothesis (RH1) concerning the improvement of the coefficient of determination cannot be directly verified quantitatively due to the lack of appropriate data (e.g. incremental R-squared).

A deeper look at the collected data allowed a comparison of the means for adjusted R-squared for three types of models (i.e. single-factor, medium-factor, and multifactor divided into the FF three-factor model and Carhart four-factor model) was made. Table 9 presents the results of such an analysis. Note that the number of studies analyzed in the table is lower than those in the qualitative analysis sections since not all the research published R-squared for their model. The differences between single-factor and medium complex seem negligible, however, the average of R-squared only seems to be higher for multifactor. To compare the R-squared means between single-factor and multifactor models and between medium complex and multifactor models, t-tests were performed. Both tests gave a p-value above 10%, i.e. insignificant difference. However, the results could be insignificant due to the small sample of papers. Therefore, one should be careful with interpreting those results.

**Table 9:** Means of adjusted R-squared for single-factor, medium complex, and multifactor models (divided into the FF three-factor model and the Carhart four-factor model).

| Models | No. of studies | Avg. R-squared | Std Dev of R-squared | Avg. / Std Dev |
|---|---|---|---|---|
| Single-factor | 11 | 0.23 | 0.29 | 0.78 |
| Medium complex | 14 | 0.20 | 0.23 | 0.85 |
| Multifactor | 13 | 0.32 | 0.26 | 1.20 |
| *FF three-factor* | *4* | *0.35* | *0.20* | *1.76* |
| *Carhart four-factor* | *9* | *0.30* | *0.30* | *1,01* |
| All | 38 | 0.24 | 0.29 | 0.81 |

*Source: The data in the table has been prepared based on articles specified in detail in the bibliography.*

Ultimately, to test the second hypothesis (RH2) that the models using more complex measures of sentiment have better predictive power, the number of research supporting this statement was analyzed. Unfortunately, only nine studies were comparing the measures of sentiment, while the analysis remained eight because one study analyzed only simple indicators. Of the remaining ones, five showed the superiority of composite indices, while three were not. This is a difference in favor of complex measures of sentiment, but it is not an unequivocal result. Therefore, this hypothesis cannot be verified.

## 5. Conclusions

First, the study provides a comprehensive review of 71 empirical papers published between 2000 and 2021, offering a broad synthesis of the literature on investor sentiment and asset pricing models. This review helps to consolidate existing knowledge by highlighting the diversity in measures of sentiment and their respective impacts on stock returns. By categorizing sentiment measures into direct and indirect types, the study provides valuable insights into their advantages and disadvantages. Direct measures, such as surveys capturing investor opinions, contrast with indirect measures derived from economic variables and market data. This detailed categorization deepens our understanding of how different sentiment measures influence asset pricing models.

The study managed to answer the research question about the impact of investor sentiment on stocks and indices returns in the presence of other market factors. The impact of sentiment was significant regardless of what variables were controlled in the research and the frequencies of the data used. These could be macroeconomic or noneconomic variables, previous returns, and factors such as SMB, HML, or WML. The study demonstrated the stability of the results over time, i.e. the results were significant regardless of the date of the study, the range of the sample period, the frequency of the data, and the division of the study period into sub-periods. However, in some cases, the results were or were not significant depending on the dependent variable, the study period, the measure used (some such as the BW index seem to be losing importance in recent years), and the asset under study.

Moreover, the research confirmed that sentiment conditioned some commonly known phenomena, such as the size premium or momentum effect. The study also showed the prevalence of the reversal effect of the sentiment, which means that the impact of sentiment on returns usually was reversed with the same magnitude after 4 or 5 days. However, the exact influence of sentiment varied greatly due to the measures used. The average impact is difficult to estimate as its direction was positive as often as negative.

The research failed to reject (both quantitatively and qualitatively) one out of the two hypotheses (RH1) presented in the article. For the second hypothesis (RH2), the number of studies was insufficient to conduct a broader analysis to confirm or reject the hypothesis. The collected data made it possible to condense the current knowledge from the most popular articles and to identify gaps requiring wider research, i.e. comparison of sentiment in various models (including multifactor models), a broad comparison between various measures of sentiment, and finding a universal measure for asset pricing models that will be competitive with the BW index and its variants. Further verification of the hypotheses verified in the study should consider a wider range of models and sentiment drivers.

The study challenges the traditional rational asset pricing models, such as the Fama-French models, by providing robust empirical evidence that investor sentiment significantly impacts asset prices. The review demonstrates that augmenting these models with sentiment proxies improves their explanatory power. This suggests that traditional models, which rely solely on rational factors, may be insufficient for capturing the complexities of real-world markets where investor sentiment plays a crucial role. While supporting behavioral finance theories, the study also challenges some aspects of these theories. Behavioral finance posits that psychological factors and irrational behavior drive market anomalies and price deviations. The analysis shows that while investor sentiment is a significant factor, its predictive power varies across different contexts, time periods, and asset classes. This variability indicates that current behavioral theories might overgeneralize the impact of sentiment and fail to account for the nuanced ways in which sentiment influences different markets. For instance, the study found that more complex sentiment measures do not always outperform simpler ones, challenging the assumption that more detailed behavioral data always lead to better predictions. The study highlights the importance of market context and conditions in determining the impact of investor sentiment, challenging the broad applicability of both traditional and behavioral theories. It shows that the significance of sentiment varies with market conditions, such as bull or bear markets, and specific asset characteristics. For example, certain sentiment measures were significant in predicting returns for small-cap stocks but not for large-cap stocks. This finding suggests that both traditional and behavioral models need to incorporate context-sensitive approaches to better capture the dynamics of different market environments. The analysis also points to the reversal effects of sentiment, where the influence of sentiment is often partially reversed in subsequent periods. This temporal variability challenges the static nature of traditional models and suggests that even behavioral models may need to incorporate more dynamic elements to account for these effects. The presence of significant lags in sentiment impact, as identified in several studies, indicates that both traditional and behavioral theories might need to be adapted to better capture the time-varying nature of sentiment effects.

Finally, the review revealed a few possible directions for the development of further research. Based on the results of this study, it is recommended that future research should focus on developing unified and robust sentiment indicators applicable across different markets and time periods, potentially enhancing predictive power through composite indices. Employing advanced analytical techniques such as machine learning and natural language processing can provide deeper insights by analyzing large volumes of text data, offering real-time sentiment analysis. Context-specific models should be developed to account for unique market characteristics, while integrating behavioral finance theories can offer a comprehensive understanding of investor sentiment's impact on market dynamics. Cross-market and temporal analyses are essential to discern global versus local sentiment influences, and real-time sentiment tracking using big data analytics can aid in high-frequency trading and risk management. Lastly, understanding sentiment-driven market fluctuations can inform policymakers and regulatory bodies in maintaining market stability and preventing speculative bubbles.

**Appendix A**

**Table 8.** All (71) papers analyzed in the study with its characteristic sorted by the year of publication.

| # | Author(s) | Asset(s) | Data period | Frequency | Investor sentiment measure(s) | Model(s) | Sign | Significance |
|---|---|---|---|---|---|---|---|---|
| 1 | Lee, C. M., Shleifer, A., & Thaler, R. H. (1991). | American - NYSE | July, 1956 and December, 1985 | Monthly | CEFD | Multifactor | Both | Yes |
| 2 | Chen, NF; Kan, R; Miller, MH (1993). | American - NYSE | July 1965 to December 1985 | Monthly | CEFD | Medium complex | Negative | Yes |
| 3 | Neal & Whitley (1998). | American - NYSE and AMEX | 1933-1993 | Monthly, Quarterly, 1,2,3, and 4-years | 1. CEFD, 2. the ratio of odd-lot sales to purchases, 3. the net mutual fund redemption on stock returns | Single-factor | Positive | 1, 3 – Yes  2 - No |
| 4 | Klibanoff, P., Lamont, O., & Wizman, T. A. (1998). | Various countries | January 1986 to March 1994 | Weekly | Based on media | Medium complex | - | No |
| 5 | Simon, DP; Wiggins, RA (2001). | American - S&P 500 | January 1989 to June 1999 | Daily | 1. the VIX, 2. The put-call ratio, 3. the TRIN | Multifactor | Positive | Yes |
| 6 | Fisher, K. L., & Statman, M. (2003). | Amercian - S&P 500 and NASDAQ | January 1989 to July 2002 | Monthly | CC measures | Single-factor | Negative | NASDAQ and small cap - Yes  The S&P500 - No |
| 7 | Brown, G., & Cliff, M. (2005) | Amercian - DJIA | January 1963 to December 2000 | Monthly | II survey results and CEFD, NYSE OOD, FUNDFLOW, ARMS index, IPON, IPORET | Single-factor / Multifactor | Negative | Yes |
| 8 | Baker, M. & Wurgler, J. (2006) | American - CRSP with share codes 10 and 11 | 07.1962-06.2001 | Monthly | BW | Multifactor | Both | Yes |
| 9 | Kumar & Lee (2006) | American - major US brokerage houses | January 1991 to November 1996 | Monthly | Buy-sell inbalance | Multifactor | Positive | Yes |
| 10 | Lemmon, M., & Portniaguina, E. (2006) | American - all CRSP | 1956 - 2002 | Monthly - 3,6, 12 months | 1. the University of Michigan survey of consumer sentiment, 2. the Conference Board survey of consumer confidence | Medium complex / Multifactor | Negative | Yes after 1977, and no before |

| # | Author(s) | Asset(s) | Data period | Frequency | Investor sentiment measure(s) | Model(s) | Sign | Significance |
|---|---|---|---|---|---|---|---|---|
| 11 | Cornelli, F., Goldreich, D., & Ljungqvist, A. (2006) | Various countries - IPO | November 1995 to December 2002 | Daily | Grey market indicators | Single-factor | Positive | Yes for high grey market prices, and no for low |
| 12 | Cook, D. O., Kieschnick, R., & Van Ness, R. A. (2006). | American - IPOs | January 1993 to December 2000 | Daily | Based on media | Medium complex | Positive | Yes |
| 13 | Tetlock (2007) | Amercian - DJIA | January 1984 to September 1999 | Daily | Based on media | Multifactor | Negative | Yes |
| 14 | Das, S. R., & Chen, M. Y. (2007) | American - Morgan Standley High-Tech Index | July to August 2001 | Daily | Based on media | Single-factor | - | No |
| 15 | Edmans, A., Garcia, D., & Norli, Ø. (2007). | Various countries | January 1973 to December 2004 | Daily | Sport game results: 1. Losses, 2. Wins | Single-factor | Negative | 1. Yes, 2. No |
| 16 | Banerjee, P. S., Doran, J. S., & Peterson, D. R. (2007) | American - NYSE, AMEX and NASDAQ | June 1986 to June 2005 | Daily | VIX | Multifactor | Positive | Yes except for portfolios based on low beta, low book to market value and large size |
| 17 | Kurov, A. (2008). | American - S&P 500 and Nasdaq-10 | 2002–2004 | Daily | 1. AAII sentiment index, 2. II sentiment index | Medium complex | Positive | No for Bull market. Yes for bear market |
| 18 | Chang, S. C., Chen, S. S., Chou, R. K., & Lin, Y. H. (2008). | American - NYSE | 1994-2004 | Intraday – hourly intervals | Weather variables: wind speed, snowiness, raininess and temperature | Multifactor | - | No |
| 19 | Fang, L., & Peress, J. (2009). | American - NYSE and NASDAQ | January 1, 1993 and December 31, 2002 | Monthly | Based on media | Multifactor | Positive | Yes |
| 20 | Schmeling (2009) | Various countries | Different for different countries | Monthly: 1, 6, 12, and 24 months | Consumer confidence | Medium complex | Negative | Yes for value, growth and small stocks; No for large stocks and for size premium for 12M and 24M horizons. |

| # | Author(s) | Asset(s) | Data period | Frequency | Investor sentiment measure(s) | Model(s) | Sign | Significance |
|---|---|---|---|---|---|---|---|---|
| 21 | Palomino, F., Renneboog, L., & Zhang, C. (2009). | UK - soccers cllubs listed on the LSE | 1999-2002 | Daily: up to three days | 1. goal difference, 2. win dummy variable, 3. loss dummy variable | Single-factor | positive for 1 and 2; negative for 3 | Yes |
| 22 | Dorn, D. (2009). | Germany - IPOs | August 1999 to May 2000 | Daily | 1. gross when issued purchases, 2. gross day plus 1 purchases | Medium complex | Negative | Yes |
| 23 | Kaplanski, G., & Levy, H. (2010). | American - NYSE | January 1950 to December 2007 | Daily | Aviation disasters | Medium complex | Negative | Yes for the first day, no for the second day |
| 24 | Kurov, A. (2010). | Amercian - S&P 500 | January 1990 to November 2004 | Daily | Multiple | Medium complex | Negative | No for Bull market, yes for bear market |
| 25 | Drakos, K. (2010). | Various countries | January 1994 to December 2004 | Daily | Terrorist activity | Multifactor | Negative | Yes |
| 26 | Da, Z., Engelberg, J., & Gao, P. (2011) | American - Russel 3000 | January 2004 to June 2008 | Weekly | Google SVI | Single-factor / Medium complex | Positive | Yes |
| 27 | Joseph, K., Wintoki, M. B., & Zhang, Z. (2011). | Amercian - S&P 500 volatility sorted portfolio deciles | 2005–2008 excl. 2004 | Weekly | Google SVI | Multifactor | Both | Yes: positive for Q7-10, negative Q1-5, no for Q6 |
| 28 | Bollen, J., Mao, H., & Zeng, X. (2011). | Amercian - DJIA | February 2008 to December 2008 | Daily | 1. Twitter OpinionFinder, 2. Twitter GPOMS | Machine learning | Impossible to detect | 1. No, 2. Yes |
| 29 | Stambaugh, R. F., Yu, J., & Yuan, Y. (2012) | American - NYSE, AMEX and NASDAQ | from July 1965 to December 2007 | Monthly | BW | Multifactor | Positive | Yes for 7 out of 11 anomalies |
| 30 | Moskowitz, T. J., Ooi, Y. H., & Pedersen, L. H. (2012) | Various countries | January 1965 to December 2009 | Quarterly | BW | Multifactor | - | Insignificant |
| 31 | Ben-Rephael, A., Kandel, S., & Wohl, A. (2012). | American - NYSE, AMEX and NASDAQ | January 1984 to December 2008 | Monthly | 1. the CSI index, 2. the BW index, 3. the VIX | Single-factor | Positive | 1,3 - Yes. 2 - No |
| 32 | Mian, G. M., & Sankaraguruswamy, S. (2012). | American - all CRSP | 1972-2007 | Daily | BW | Multifactor | Negative | Yes |
| 33 | Hribar, P., & McInnis, J. (2012). | American - all CRSP | August 1982 to December 2005 | Monthly | BW | Multifactor | - | No |

| # | Author(s) | Asset(s) | Data period | Frequency | Investor sentiment measure(s) | Model(s) | Sign | Significance |
|---|---|---|---|---|---|---|---|---|
| 34 | Białkowski, J., Etebari, A., & Wisniewski, T. P. (2012). | Various countries | 1989–2007 | Daily | Ramadan | Medium complex | Positive | Yes when at least 50% of citizens were Muslims |
| 35 | Chung, S. L., Hung, C. H., & Yeh, C. Y. (2012). | American NYSE, AMEX and NASDAQ | January 1966 to December 2007 | Monthly | BW | Multifactor | Positive | Yes |
| 36 | Garcia, D. (2013). | Amercian - DJIA | 1905-2005 | Daily | Media positive, negative, pessimism | Single-factor | Both: positive for 1; negative for 2 and 3 | Yes |
| 37 | Han, Y., Yang, K., & Zhou, G. (2013) | American - NYSE and AMEX | July 1965 to December 2007 | Daily | BW | Multifactor | - | No |
| 38 | Corredor, P., Ferrer, E., & Santamaria, R. (2013). | Various countries | 1990-2007 | Monthly: 6, 12, 24 Months | 1. BW, 2. EU sentiment measure | Multifactor | Both depending on portfolio | 1 – Yes, 2 - No |
| 39 | Xiong, G., & Bharadwaj, S. (2013). | American - all CRSP, Ken French's website and Compustat | November 2004 to February 2010 | Monthly | Based on media: positive and negative frequencies of news | Multifactor | Both | Both |
| 40 | Yu, Y; Duan, WJ; Cao, Q (2013). | American - CRSP and compustat | July 2011 to September 2011 | Daily | Media: 1. blogs, 2. forums, 3. Twitter, 4. news web page | Multifactor | Both : positive for blog, and negative for forum | 2,3 -Yes, 1,4 - No |
| 41 | Chen, H., De, P., Hu, Y. J., & Hwang, B. H. (2014). | American - all CRSP | 2005 - 2012 | Daily | 1. freq. of negative words in articles of SA, 2 freq. of comments in SA, 3. the average fraction of negative words in DJNS | Single-factor | Negative | 1,2 – Yes, 3 - No |
| 42 | Sprenger, T. O., Tumasjan, A., Sandner, P. G., & Welpe, I. M. (2014). | Amercian - S&P 100 | January 2010 to June 2010 | Daily | Twitter | Single-factor | Positive | Yes |
| 43 | Li, Q., Wang, T., Li, P., Liu, L., Gong, Q., & Chen, Y. (2014). | Chinese - CSI 100 | 2011 | Daily | Based on media | machine learning | Impossible to detect | Yes |

| # | Author(s) | Asset(s) | Data period | Frequency | Investor sentiment measure(s) | Model(s) | Sign | Significance |
|---|---|---|---|---|---|---|---|---|
| 44 | Kim, S. H., & Kim, D. (2014). | American - 91 firms posted on theYahoo! Finance message board | January 2005 to December 2010 | Monthly, weekly, daily | Based on media | Single-factor | Positive | Yes |
| 45 | Siganos, A., Vagenas-Nanos, E., & Verwijmeren, P. (2014). | Various countries | September 2007 to March 2012 | Daily | Facebook's Gross National Happiness Index | Single-factor | Positive | Yes |
| 46 | Hillert, A., Jacobs, H., & Müller, S. (2014). | American - NYSE, AMEX and NASDAQ: winners, losers, and mid returns | January 1989 to December 2010 | Monthly | Based on media | Multifactor | Positive | Yes only for winners |
| 47 | Takeda, F., & Wakao, T. (2014). | Japanese - Nikkei 225 | January 2008 to December 2011 | Weekly | Google SVI | Multifactor | Positive | Yes |
| 48 | Fong, W. M., & Toh, B. (2014). | American - NYSE, AMEX and NASDAQ | July 1965 to December 2007 | Monthly | BW | Multifactor | Negative | Yes |
| 49 | Li, XD; Xie, HR; Chen, L; Wang, JP; Deng, XT (2014). | Chinese - Hong Kong stock exchange | January 2003 to March 2008 | Daily | Based on media | Machine learning | Impossible to detect | Yes |
| 50 | Da, Z., Engelberg, J., & Gao, P. (2015). | Amercian - SP500, NASDAQ, Russel 1000 | January 2004 to December 2011 | Daily | Google SVI | Medium complex | Negative | Yes |
| 51 | Huang, D., Jiang, F., Tu, J., & Zhou, G. (2015). | Amercian - S&P 500 | July 1965 to December 2010 | Monthly | 1. the BW index, 2. The BW based on the PLS procedure | Single-factor | Negative | 1 – No, 2 - Yes |
| 52 | Stambaugh, R. F., Yu, J., & Yuan, Y. (2015) | American - NYSE, AMEX and NASDAQ | August 1965 to January 2011 | Monthly | BW | Multifactor | Negative | Yes |
| 53 | Goetzmann, W. N., Kim, D., Kumar, A., & Wang, Q. (2015). | American - all CRSP | January 1999 to December 2010 | Monthly | The sky cloud cover variables | Medium complex | Positive | Yes |
| 54 | Jacobs, H. (2015). | American - CRSP and compustat | Different periods | Monthly | BW | Multifactor | Both | Yes for most anomalies |
| 55 | Ni, Z. X., Wang, D. Z., & Xue, W. J. (2015). | Chinese - Shanghai Stock Exchange (SSE) Large & Mid & Small Cap Index | January 2005 to September 2013 | Monthly: 1-24 months | Opening accounts number and turnover rate | Multifactor | Positive for stocks with high returns in the short term and negative in the long term | Yes |

| # | Author(s) | Asset(s) | Data period | Frequency | Investor sentiment measure(s) | Model(s) | Sign | Significance |
|---|---|---|---|---|---|---|---|---|
| 56 | Ding, X; Zhang, Y; Liu, T; Duan, JW (2015). | American - S&P 500 | October 2006 to November 2013 | Daily | Based on media | Machine learning | Impossible to detect | Yes |
| 57 | Nguyen, TH; Shirai, K; Velcin, J (2015). | American | July 2012 to July 2013 | Daily | Based on media | Machine learning | Impossible to detect | Yes |
| 58 | Ranco, G; Aleksovski, D; Caldarelli, G; Grcar, M; Mozetic, I (2015). | Amercian - DJIA | May 2013 to September 2014 | Daily | Twitter | Machine learning | Positive | Yes |
| 59 | Bijl, L., Kringhaug, G., Molnár, P., & Sandvik, E. (2016). | Amercian - S&P 500 | January 2007 to December 2013 (subprime crisis, i.e. 2008-2013, and period after crisis, i.e. 2010-2013. | Monthly | Google SVI | Medium complex | Both | Yes |
| 60 | Antoniou, C., Doukas, J. A., & Subrahmanyam, A. (2016). | American - NYSE, AMEX and NASDAQ | 1966-2010 | Monthly | BW | Medium complex | Positive | Yes |
| 61 | Stambaugh, R. F., & Yuan, Y. (2017). | American - NYSE, AMEX and NASDAQ | January 1967 to December 2013 | Monthly | BW | Multifactor | For MGMT and PERF, negative on short legs, and positive for long-short. | For MGMT and PERF significant For market and SMB insignificant. |
| 62 | Oliveira, N., Cortez, P., & Areal, N. (2017). | American - S&P 500, RUSELL 2000, DJIA, NASDAQ 100 | January 2014 to June 2014 | Daily | Multiple | Machine learning | Impossible to detect | Yes |
| 63 | You, W., Guo, Y., Zhu, H., & Tang, Y. (2017). | Chinese - the Resset Financial Database China's industry data | January 1995 to March 2016 | Monthly | EPU | Medium complex | Negative | Yes |
| 64 | Chen, J., Jiang, F., & Tong, G. (2017). | Chinese - All A-share stocks listed in Shanghai and Shenzhen stock exchanges | January 1996 to December 2013 | Monthly | EPU | Medium complex | Negative | Yes |
| 65 | Renault, T. (2017). | American - S&P 500, DJIA and NASDAQ | January 2012 to December 2016 | Intraday at half-hour intervals | Based on media | Single-factor | Positive | Yes |

| # | Author(s) | Asset(s) | Data period | Frequency | Investor sentiment measure(s) | Model(s) | Sign | Significance |
|---|---|---|---|---|---|---|---|---|
| 66 | Ichev, R., & Marinč, M. (2018). | American NYSE and NASDAQ | January 2014 to June 2016 | Daily | Ebola outbreak events | Medium complex | Negative | Yes |
| 67 | Bartov, E., Faurel, L., & Mohanram, P. S. (2018). | American - Russel 3000 | January 2009 to December 2012 | Daily | Twitter | Multifactor | Positive | Yes |
| 68 | Phan, D. H. B., Sharma, S. S., & Tran, V. T. (2018). | Various countries | Different periods for different countries | Monthly | 1. Global EPU, 2. Local EPU | Multifactor | Negative | Both |
| 69 | Weng, B; Lu, L; Wang, X; Megahed, FM; Martinez, W (2018). | American | 2013 - 2016 | Daily | Wikipedia hits, financial news, Google trends and technical indicators | Machine learning | Impossible to detect | Yes |
| 70 | Jiang, F., Lee, J., Martin, X., & Zhou, G. (2019). | American - all CRSP, Ken French's website and Compustat | January 2003 to December 2014 | Monthly | 1. the BW index, 2. the Huang et al. (2015) investor sentiment index, 3. the CSI, 4. the Conference Board Consumer Confidence index, 5. FEARS | Single-factor / Medium complex | Negative | Yes |
| 71 | Liu, H., Manzoor, A., Wang, C., Zhang, L., & Manzoor, Z. (2020). | Various countries | January 2020 to March 2020 | Daily | COVID-19 cases | Single-factor | Negative | Yes |

*Source: The data in the table has been prepared based on articles specified in detail in the bibliography.*